\documentclass[pra,twocolumn,aps,showpacs,nofootinbib,floatfix,superscriptaddress]{revtex4-1}
\usepackage{amsmath}
\usepackage{amssymb}
\usepackage{graphicx}
\usepackage{verbatim}
\usepackage[colorlinks=true,linkcolor=blue,citecolor=blue,urlcolor=blue]{hyperref}
\usepackage{txfonts}

\begin{document}

\title{Nonlinear Spectroscopy of Trapped Ions}
\author{Frank Schlawin}
\email{frank.schlawin@physik.uni-freiburg.de}
\affiliation{Physikalisches Institut, Albert-Ludwigs-Universit\"at Freiburg, Hermann-Herder-Stra\ss e 3, 79104 Freiburg, Germany}
\affiliation{Department of Chemistry, University of California, Irvine, California 92697, USA}
\author{Manuel Gessner}
\email{manuel.gessner@physik.uni-freiburg.de}
\affiliation{Physikalisches Institut, Albert-Ludwigs-Universit\"at Freiburg, Hermann-Herder-Stra\ss e 3, 79104 Freiburg, Germany}
\affiliation{Department of Physics, University of California, Berkeley, California 94720, USA}
\author{Shaul Mukamel}
\affiliation{Department of Chemistry, University of California, Irvine, California 92697, USA}
\author{Andreas Buchleitner}
\affiliation{Physikalisches Institut, Albert-Ludwigs-Universit\"at Freiburg, Hermann-Herder-Stra\ss e 3, 79104 Freiburg, Germany}
\date{\today}
\pacs{78.47.jh, 37.10.Ty, 78.47.da, 05.60.Gg}

\begin{abstract}
Nonlinear spectroscopy employs a series of laser pulses to interrogate dynamics in large interacting many-body systems, and has become a highly successful method for experiments in chemical physics. Current quantum optical experiments approach system sizes and levels of complexity which require the development of efficient techniques to assess spectral and dynamical features with scalable experimental overhead. However, established methods from optical spectroscopy of macroscopic ensembles cannot be applied straightforwardly to few-atom systems. Based on the ideas proposed in [M. Gessner \textit{et al.} \href{http://dx.doi.org/10.1088/1367-2630/16/9/092001}{New J. Phys. \textbf{16} 092001 (2014)}], we develop a diagrammatic approach to construct nonlinear measurement protocols for controlled quantum systems and discuss experimental implementations with trapped ion technology in detail. These methods in combination with distinct features of ultra-cold matter systems allow us to monitor and analyze excitation dynamics in both the electronic and vibrational degrees of freedom. They are {\em independent} of system size, and can therefore reliably probe systems where, e.g., quantum state tomography becomes prohibitively expensive. We propose signals that can probe steady state currents, detect the influence of anharmonicities on phonon transport, and identify signatures of chaotic dynamics near a quantum phase transition in an Ising-type spin chain.
\end{abstract}

\maketitle

\section{Introduction}

Nonlinear spectroscopy has proven to be an indispensable tool for the analysis of many-body dynamics and transport phenomena in complex systems \cite{Ernst, Haeberlen, HammZanni, Shaul_book,Shaul-accchem, Kauffmann}. Amongst many features, it can provide information on many-body interactions beyond mean-field theory \cite{CundiffPRL2}, collective resonances \cite{CundiffPRL}, as well as environment-induced energy transport pathways \cite{Brixner}. Experiments are often carried out under extremely challenging conditions characterized by short timescales and tight spatial confinement. For instance, in vividly debated experiments on photosynthetic complexes \cite{Brixner,Engel}, typical timescales for energy transport and coherence decay are on the order of picoseconds. Interacting chromophores are separated only by few nanometers, two orders of magnitude below the diffraction limit of optical light. These parameters render precise experimental control of such molecular aggregates extremely demanding.

Quantum information applications in contrast are characterized by flexibility in system engineering and low decoherence rates \cite{Leibfried,HARTMUTREVIEW,CHRISTIAN,ColdAtoms,RydbergReview,Schoelkopf}. For instance, the electronic degrees of freedom of individual ions in a linear trap can be controlled experimentally with high spatial and temporal precision using focussed lasers \cite{HARTMUTREVIEW}. The potential to mimic a vast range of closed and open systems renders this technology potentially valuable for simulations of complex quantum models \cite{CHRISTIAN,Julio,Schaetz,AspuruQC} on much longer timescales. Typically, trapped ion experiments take place on the order of microseconds while coherence can be upheld for milliseconds. Furthermore, the creation and observation of spatially localized vibrational excitations has recently been achieved experimentally \cite{Haze,Monroe13,Harlander, Brown}. This opens up new opportunities \cite{Plenio13,Mike} for the study of non-equilibrium phenomena, such as energy transport on smallest scales and under well-controlled conditions. For the efficient assessment of such phenomena one cannot resort to established methods which are restricted to probe excitations in thermal equilibrium \cite{Leibfried}. Alternative approaches from quantum information theory, such as quantum state or process tomography are neither practical nor feasible in increasingly complex quantum systems. Thus, scalable methods need to be developed for the systematic analysis of the dynamics of large controllable many-body systems. The adaption of methods from nonlinear spectroscopy promises to be beneficial \cite{Letter}: Specifically designed pulse sequences that are independent of the system size provide a powerful method for probing multi-point correlation functions, which in turn reveal spectral and dynamical properties of the system. We address the following two questions in this paper:
\begin{itemize}
\item First, how can methods from nonlinear spectroscopy be applied to trapped ion systems and which additional information can be obtained thereby? 
\item Second, how can single-site addressability in neutral atoms \cite{Meschede,Markus10,Weitenberg11} or ions \cite{HARTMUTREVIEW} be used to extend standard nonlinear spectroscopy techniques?
\end{itemize}

\begin{figure}[t]
\centering
\includegraphics[width=.49\textwidth]{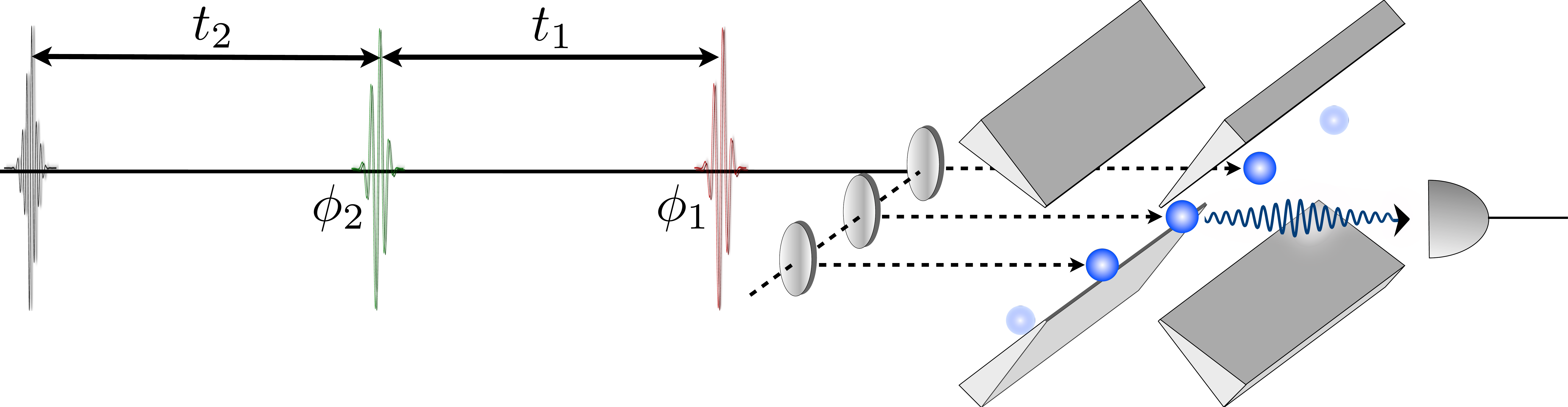}
\caption{(Color online) Experimental setup for a second-order nonlinear signal, which can monitor coherent and incoherent excitation transport, quantum phase transitions, or steady state currents. Two focussed laser pulses with controlled phases $\phi_1$ and $\phi_2$ excite individual ions in a linear trap. A readout pulse induces fluorescence which is collected to measure the nonlinear response of the system. The two time delays $t_1$ and $t_2$ are scanned. The phases of the excitation pulses are used to distinguish between various quantum pathways.}
\label{fig.1}
\end{figure}

The basic idea of a nonlinear spectroscopic experiment in a quantum optical setting is depicted in Fig.~\ref{fig.1}: A linear chain of trapped ions is locally excited and probed by a series of phase-coherent pulses, and the time delays between these pulses are scanned. Following the interaction with the pulse sequence, the fluorescence signal is collected. Both, electronic and vibrational degrees of freedom can be investigated in a Coulomb crystal of trapped ions. The motions of different positively charged ions in a common trap potential are coupled by long-range Coulomb repulsion. The electronic states of individual ions are only negligibly coupled, but, using the common motional modes as a coupling agent, effective spin-spin interactions may be engineered with appropriate laser fields \cite{PorrasCiracSpins}. 

This paper is structured as follows: In section \ref{theory}, we present a general diagrammatic approach to nonlinear spectroscopic measurements. Sections \ref{electronic-DOF} and \ref{vibrational-DOF} discuss the phononic and electronic degrees of freedom for a linear chain of trapped ions, and specify experimental implementations of excitation and readout protocols. In sections \ref{sec.sqc} to \ref{sec.photon-echo}, we present numerical simulations \cite{qutip} for selected applications.
\section{Diagrammatic construction of multidimensional measurement protocols}
\label{theory}
In this section we introduce a diagrammatic representation of nonlinear measurement protocols. The presentation is intentionally kept general. Later in this manuscript, we will present applications to electronic and vibrational degrees of freedom of trapped ions. However, it must be stressed that our formalism may also be applied to similar synthetic quantum matter systems such as ultracold atoms in optical lattices \cite{ColdAtoms} or trapped Rydberg atoms, e.g., in optical tweezers \cite{Tweezer}.

\subsection{Interaction, time evolution and readout}
The first basic building block for spectroscopic signals is the interaction of the probe with a controllable external field. This interaction can induce transitions in the system, leading to the creation or destruction of excitations. In the following, we only consider pulses whose duration is much shorter than the dynamics of the sample system, thus they can be approximated as impulsive interactions \cite{Shaul_book}. This allows us to avoid the explicit treatment of the optical fields, and to describe the evolution during the interaction with the probe with a transition operator acting solely in matter space:
\begin{align}\label{eq.generalint}
V_i=\alpha\mathbb{I}+\beta e^{i\phi_i}A_i^{\dagger}+\gamma e^{-i\phi_i}A_i,
\end{align}
where $A_i$ ($A_i^{\dagger}$) deexcites (excites) a mode labeled by $i$. The phase $\phi_i$ is a tunable parameter of the interaction field, and the real amplitudes $\alpha$, $\beta$ and $\gamma$ depend on intensity and duration of the pulse. 

Before we proceed to the construction of multidimensional signals, we briefly discuss examples of such interaction operators. As we will discuss in detail in Sec.~\ref{electronic-DOF}, such excitations may be induced in the electronic degree of freedom of individual ions with strong resonant pulses, leading to $\beta=\gamma$, whereas $\alpha$ and $\beta$ can be tuned arbitrarily under the constraint $\alpha^2+\beta^2=1$. In this case $A_i=\sigma^{(i)}_-$, where $\sigma^{(i)}_{\pm}=1/2(\sigma^{(i)}_x\pm i\sigma^{(i)}_y)$ and $\sigma^{(i)}_i$ denote the Pauli matrices for ion $i$, whose electronic state may be conceived as an effective spin-$1/2$ system. In Sec.~\ref{sec.phonons} we discuss the excitation of local phonons, induced, for instance, by weak kicks described by $\alpha\approx1$, and $\beta=-\gamma\ll1$. We then have $A_i=a_i$ destroying one phonon at ion $i$. Alternatively, by design, one may only address transitions which excite or destroy local phonons, such that $\beta=0$ or $\gamma=0$, see Sec.~\ref{sec.selectiveexcitation}. Finally we note that by focussing the interaction laser on more than one ion -- or even the entire chain -- one may excite various local phonons $A_i=a_{i_1}+a_{i_2}+\dots$ or collective vibrations $A_i=b_i$, where $b_i$ is the destruction operator for a vibrational eigenmode of all ions.

\begin{figure}[tb]
\centering
\includegraphics[width=.4\textwidth]{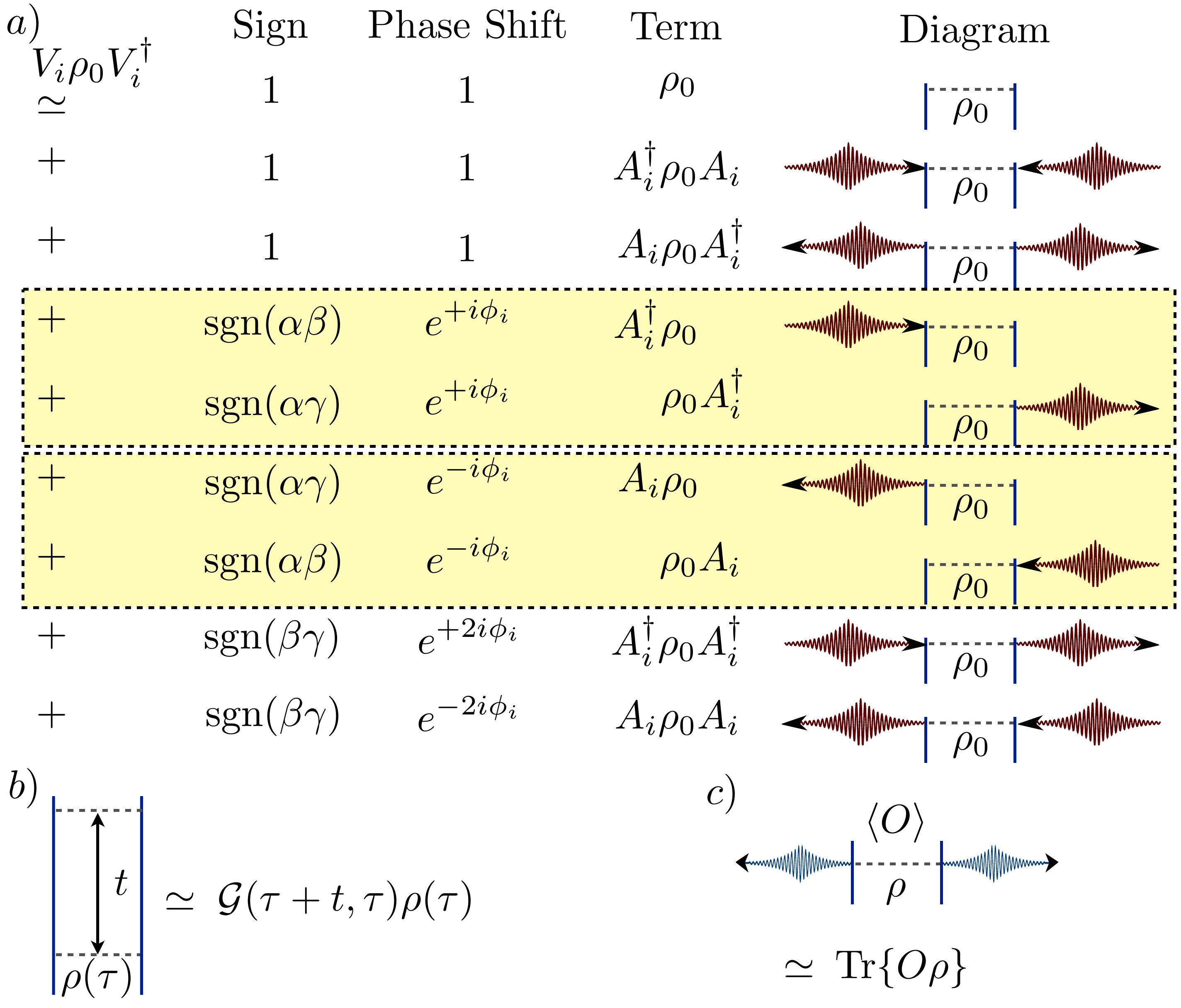}
\caption{(Color online) Elementary diagrams for the systematic construction of nonlinear measurement protocols, based on the interaction operator in Eq.~(\ref{eq.generalint}). a) Excitation and de-excitation events describing an interaction with an external field. The two sides of the ladder represent \textit{ket} and \textit{bra} side of the density matrix. Arrows pointing right (left) denote the action of $A^{\dagger}_i$ ($A_i$) and carry a phase shift of $e^{+i\phi_i}$ ($e^{-i\phi_i}$). On the \textit{bra} side the roles of $A_i$ and $A_i^{\dagger}$ are exchanged. Each term carries a specific sign (sgn) which depends on the form of the interaction operator. The highlighted terms describe coherences of the density matrix and may be extracted by phase cycling. b) Time evolution in absence of interactions with the interaction field is described by the Green's function $\mathcal{G}$. In diagrammatic representation, the time axis is chosen upwards. c) Readout of the observable $O$ is represented by two blue arrows.}
\label{fig.elementarydiagrams}
\end{figure}

A general transition operator $V_i$, as given in Eq.~(\ref{eq.generalint}), transforms an initial state $\rho_0=\rho(\tau_0)$ according to the superoperator $\mathcal{V}_i \rho_0 = V_{i}\rho_0V^{\dagger}_{i}$ whose action yields
\begin{align}\label{eq.interact}
\mathcal{V}_i \rho_0 
%=\:&(\alpha\mathbb{I}+\beta e^{i\phi_{i}}A_{i}^{\dagger}+\gamma e^{-i\phi_{i}}A_{i})\rho_0(\alpha^*\mathbb{I}+\beta^* e^{-i\phi_{i}}A_{i}+\gamma^* e^{i\phi_{i}}A^{\dagger}_{i})\notag\\
=\:&\alpha^2\rho_0+\beta^2A_{i}^{\dagger}\rho_0A_{i}+\gamma^2A_{i} \rho_0A^{\dagger}_{i}\notag\\
&+\alpha e^{-i\phi_{i}}\left(\beta \rho_0A_{i}+\gamma A_{i}\rho_0\right)\notag\\&+\alpha e^{i\phi_{i}}\left(\gamma\rho_0A^{\dagger}_{i}+\beta A_{i}^{\dagger}\rho_0\right)\notag\\&+e^{-2i\phi_{i}}\gamma\beta A_{i}\rho_0A_{i}+e^{2i\phi_{i}}\beta\gamma A_{i}^{\dagger}\rho_0A_{i}^{\dagger}.
\end{align}
Representing an excitation ($A_{i}^{\dagger}$) on the \textit{ket} side of the density matrix by an arrow pointing right, and a de-excitation $(A_i)$ by an arrow pointing left, we can depict all of the contributing terms by the elementary diagrams shown in Fig.~\ref{fig.elementarydiagrams} a). Note that, if $V_i$ represents a linear approximation in terms of the interaction strength (e.g. an expansion of a displacement operator), not all quadratic terms are present in Eq.~(\ref{eq.interact}). Under these conditions, it is sufficient to keep only the linear contributions in Eq.~(\ref{eq.interact}), corresponding to those diagrams with only a single arrow in Fig.~\ref{fig.elementarydiagrams} a). We can group the contributing terms according to the phase shift which is imprinted onto the quantum state. This phase shift is crucial for nonlinear spectroscopy since it allows to extract the contribution of a particular set of quantum pathways by a technique known as phase cycling \cite{Ernst}. We next discuss another important ingredient of nonlinear protocols: the free time evolution between interactions.

To construct a nonlinear measurement protocol, we design sequences of pulses, separated by tunable time-intervals where the system evolves freely without interactions with the external field. The time evolution is governed by the system Hamiltonian and possibly by the influence of an environment. Formally the evolution from time $\tau_a$ to $\tau_b$ is described by a Green's function $\mathcal{G}(\tau_b,\tau_a)=\mathcal{T}\exp(-i\int_{\tau_a}^{\tau_b}dt\mathcal{L}(t))$, generated by a Liouvillian $\mathcal{L}(t)$, and $\mathcal{T}$ denotes the time ordering operator. Here, we only consider time-independent Liouvillians, such that we have $\mathcal{G}(\tau_b-\tau_a)=\exp[-i\mathcal{L}(\tau_b-\tau_a)]$. In the diagrammatic theory, the time axis is chosen upwards, see Fig.~\ref{fig.elementarydiagrams} b). Finally, at the end of the sequence, a measurement of an observable $O_j$ is carried out at ion $j$, represented by two outgoing blue arrows in the diagram, see Fig.~\ref{fig.elementarydiagrams} c).

As a first example, we consider the time evolution of the initial density matrix $\rho_0=\rho(\tau_0)$ under a sequence of two pulses,
\begin{align}
\rho^{(2)}_{i_1,i_2}(t_1,t_2)=\mathcal{G}(t_2)\mathcal{V}_{i_2}\mathcal{G}(t_1)\mathcal{V}_{i_1}\rho(\tau_0),
\end{align}
where we have introduced the time intervals $t_i=\tau_i-\tau_{i-1}$. Upon readout, we obtain the second-order signal
\begin{align}\label{eq.2ndorder}
S^{(2)}_{i_1,i_2;j}(t_1,t_2)=\mathrm{Tr}\{O_j\rho^{(2)}_{i_1,i_2}(t_1,t_2)\}.
\end{align}
In order to extract the contribution of a particular quantum pathway from $S^{(2)}_{i_1,i_2;j}(t_1,t_2)$, we can employ phase cycling.

\begin{figure}[tb]
\centering
\includegraphics[width=.5\textwidth]{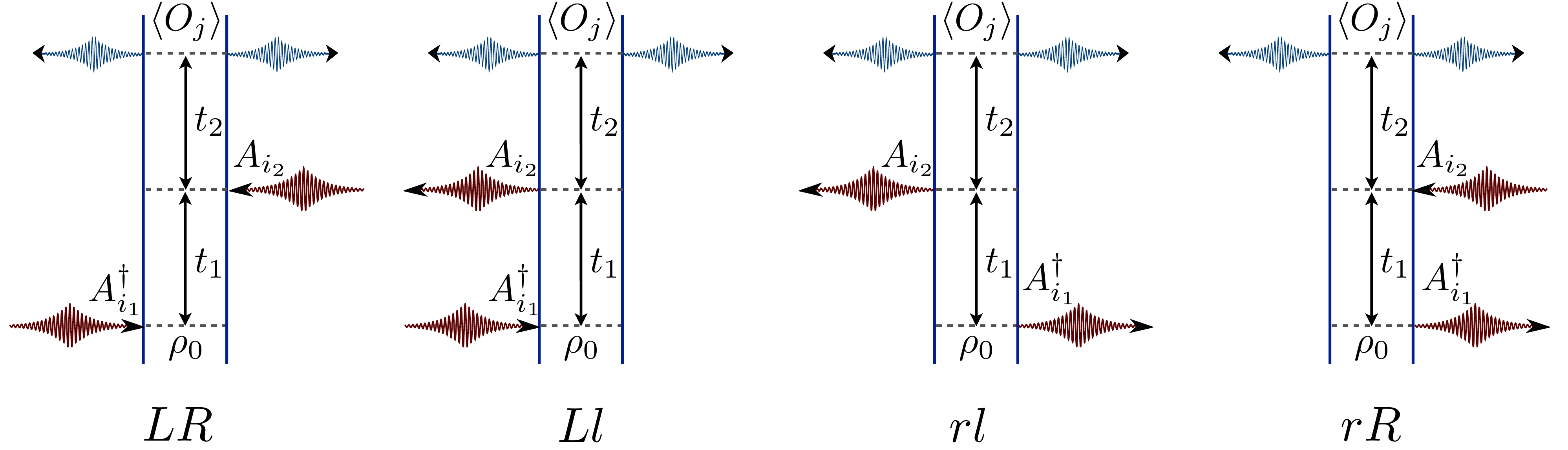}
\caption{(Color online) All second order diagrams with phase signature $\phi_{i_1}-\phi_{i_2}$ comprising the single quantum coherence (\textit{SQC}) signal. If the initial state does not contain excitations, $\rho_0=|0\rangle\langle 0|$, and the observable $O_j$ detects excitations, $\langle 0|O_j|0\rangle=0$, only the leftmost diagram yields a non-vanishing signal. We denote each diagram using capital (small) letters for excitations (de-excitations) and L/l (R/r) for the left (right) side of the diagram.}
\label{fig.sqc}
\end{figure}

\subsection{Phase cycling}
As we have seen in the previous section, different excitations may be distinguished via the phase factor which is imprinted onto the quantum state by the driving field. In nonlinear spectroscopy of molecules in the bulk, the final pulse stimulates emission with a well-defined wave vector, which may be used to extract certain pathways by spatial phase-matching \cite{Shaul_book}. In trapped-ion experiments, fluorescence is induced by an additional readout pulse and occurs into random spatial directions. It is thus not possible to use the phase matching technique. Hence, in order to separate these contributions individually in terms of their phase shift, we employ the phase cycling protocol, which originally was introduced in the context of nuclear magnetic resonance \cite{Ernst}, but has also been implemented in optical signals \cite{Warren} and even single molecule spectroscopy \cite{Hulst,LottPNAS}.

Phase cycling relies on the ability to control the phases $\phi_i$ of individual pulses. The basic principle is to scan the phases over a well-chosen set of values, such that a discrete Fourier transform of the measured data is able to filter out the desired dependence on these phases. First note that the total signal contains only terms which depend on the phase difference between the applied pulses, since the readout is only nonzero if the pathway leads to a measureable population at the end of the pulse sequence. This means that for a two-pulse sequence only the phase dependences $n(\phi_{i_1}-\phi_{i_2})$, $n=0,\pm1,\dots,\pm n_{\mathrm{max}}$ contribute, where $n_{\mathrm{max}}$ depends on the form of the interaction. We first group the final signal into terms with the same dependence on $\Delta\phi=\phi_{i_1}-\phi_{i_2}$ as 
\begin{align}
S^{(2)}(\Delta\phi)=\sum_{n=-n_{\mathrm{max}}}^{n_{\mathrm{max}}}S^{(2)}_ne^{in\Delta\phi}.
\end{align}
In order to extract the complex-valued terms $S^{(2)}_n$ we employ an inverse discrete Fourier transform:
\begin{align}
S^{(2)}_n=\frac{1}{2n_{\mathrm{max}}+1}\sum_{k=0}^{2n_{\mathrm{max}}}S^{(2)}(\delta\phi_k)e^{-i n\delta\phi_k},
\end{align}
with $\delta\phi_k=2\pi k/(2n_{\mathrm{max}}+1)$. Experimentally this means that individual terms with distinct phase dependences can be extracted by scanning $\Delta\phi$ over the values $2\pi k/(2n_{\mathrm{max}}+1)$, $k\in[0,2n_{\mathrm{max}}]$ followed by post-processing the obtained data sets $S^{(2)}(\delta\phi_k)$ with a discrete Fourier transform.

Coming back to the signal derived in Eq.~(\ref{eq.2ndorder}), our goal is to extract those terms which have acquired the phase shift $e^{i(\phi_{i_1}-\phi_{i_2})}$, see the yellow highlighted elements in Fig.~\ref{fig.elementarydiagrams} a). This leads to the single quantum coherence (\textit{SQC}) signal \cite{Letter}, represented by the diagrams in Fig.~\ref{fig.sqc}. Each diagram represents one quantum pathway. Generally, all diagrams must be added up to obtain the total signal. It is important to keep track of their relative signs, which depend on the interaction, as shown in Fig.~\ref{fig.elementarydiagrams} a). We can explicitly translate these diagrams into signals, leading to
\begin{align}\label{eq.totalsqc}
S^{(\textit{SQC})}_{ i_1, i_2; j}(t_1,t_2)=\:&S^{(\textit{LR})}_{i_1, i_2; j}(t_1,t_2)+\mathrm{sgn}(\gamma\beta)S^{(\textit{Ll})}_{i_1, i_2; j}(t_1,t_2)\notag\\&+S^{(\textit{rl})}_{i_1, i_2; j}(t_1,t_2)+\mathrm{sgn}(\gamma\beta)S^{(\textit{rR})}_{i_1, i_2; j}(t_1,t_2)
\end{align}
with
\begin{align}
S^{(\textit{LR})}_{ i_1, i_2; j}(t_1,t_2)&=\mathrm{Tr}\{O_j\mathcal{G}(t_2)\mathcal{A}^{(R)}_{i_2} \mathcal{G}(t_1) \mathcal{A}^{(L)}_{i_1}\rho(0)\}\label{eq.sqcleftmost}\\
S^{(\textit{Ll})}_{ i_1, i_2; j}(t_1,t_2)&=\mathrm{Tr}\{O_j\mathcal{G}(t_2)\mathcal{A}^{(l)}_{i_2}\mathcal{G}(t_1)A^{(L)}_{i_1}\rho(0)\}\\
S^{(\textit{rl})}_{ i_1, i_2; j}(t_1,t_2)&=\mathrm{Tr}\{O_j\mathcal{G}(t_2)\mathcal{A}^{(l)}_{i_2}\mathcal{G}(t_1) \mathcal{A}^{(r)}_{i_1} \rho(0)\}\\
S^{(\textit{rR})}_{ i_1, i_2; j}(t_1,t_2)&=\mathrm{Tr}\{O_j\mathcal{G}(t_2) \mathcal{A}^{(R)}_{i_2} \mathcal{G}(t_1) \mathcal{A}^{(r)}_{i_1} \rho(0)\}.
\end{align}
Here, we have defined the Liouville space superoperators $\mathcal{A}_i^{(L)} X = A_i^{\dagger} X$, $\mathcal{A}_i^{(R)} X = X A_i$, $\mathcal{A}_i^{(l)} X = A_i X$ and $\mathcal{A}_i^{(r)} X = XA_i^{\dagger}$. If the initial state contains no excitations, all pathways which involve a de-excitation of the initial state vanish, and in Fig.~\ref{fig.sqc} only the two leftmost contributions $S^{(\textit{LR})}$ and $S^{(\textit{Ll})}$ remain. If the observable $O_j$ measures excitations in the system, there will be no signal if the system is in the ground state at the time of the readout. This leaves only the pathway on the far left, $S^{(\textit{LR})}$. This pathway represents the contribution of a coherence between the ground state and a single excited state which evolves during $t_1$. Then, a second interaction excites the system in a coherence of single excited states, evolving during $t_2$. Note that the two excitations, as well as the readout pulse may be induced at different sites in a chain of ions, which provides a large range of possibilities to probe the system with spatial resolution. Applications for the \textit{SQC} signal are discussed in Sec.~\ref{sec.sqc}.

For transitions of the form of Eq.~(\ref{eq.generalint}), we have $n_{\mathrm{max}}=2$ (cf. Fig.~\ref{fig.elementarydiagrams}), and if the initial state is the ground state, $\rho_0=|0\rangle\langle0|$, we obtain $n_{\mathrm{max}}=1$. Thus $2n_{\mathrm{max}}+1=3$ repetitions of the experiment are sufficient to extract the \textit{SQC} signal $S^{(2)}_1$ in this case. Larger values of $n_{\mathrm{max}}$ are possible for stronger interactions or sequences which contain several pulses with identical phases, effectively leading to terms of the form $(A_i^{\dagger})^me^{im\phi_i}$.%, see Sec.~\ref{sec.multiexciton}.

For higher-order signals, a phase difference $\Delta\phi_i$ appears between each pair of consecutive pulses. The dependence on these can be extracted as above using multidimensional inverse Fourier transforms. For further details on phase cycling we refer to Ref.~\cite{Ernst}.

This diagrammatic approach represents a convenient method for the systematic construction of multidimensional measurement protocols and is easily extended to higher order signals. In general, a sequence of $n$ pulses creates the nonequilibrium density matrix
\begin{align}
\rho_{i_1,\dots,i_n}^{(n)}(t_1,\dots,t_n)= \prod_{k=1}^n\left[\mathcal{G}(\tau_k,\tau_{k-1})\mathcal{V}_{i_k}\right]\rho(\tau_0), \label{rho^n}
\end{align}
which is probed via the $n$-th order signal
\begin{align}
S^{(n)}_{i_1, \dots, i_n; j} (t_1,\dots,t_n) &= \mathrm{Tr} \{ O_j \rho_{i_1,\dots,i_n}^{(n)}(t_1,\dots,t_n) \},
\end{align}
again may be dissected into groups of quantum pathways using phase cycling. These signals may be regarded as generalizations of Ramsey fringes \cite{Ramsey}, which have been considered recently to probe many body observables in thermal equilibrium of quantum systems \cite{Lukin13}.

%\subsection{Diagram rules}
%A set of rules for the construction of arbitrary signals can be given based on elementary diagrams shown in Fig.~(\ref{fig.elementarydiagrams}):
%\begin{itemize}
%	\item[R1:] An excitation (de-excitation) is represented by an arrow pointing right (left) on the \textit{ket} side, and vice versa on the \textit{bra}.
%	\item[R2:] An arrow pointing right (left) yields a phase-shift of $+ \phi_i$ ($- \phi_i$).
%	\item[R3:] Each interaction carries a sign which eventually determines whether two diagrams with the same phase dependence are added or subtracted to obtain the final signal.
%	\item[R4:] The readout pulse is represented by de-excitations on both sides of the density matrix. It only yields a nonvanishing contribution, if the pathway had an identical number of excitations on both \textit{ket} and \textit{bra} before readout.
%\end{itemize}
%Besides, we employ the rotating-wave approximation, $i.e.$ we discard diagrams which feature de-excitations of the ground state, for instance.

%Note that rule R3 is in strong contrast to regular diagram rules in nonlinear spectroscopy, where each interaction on the \textit{bra} yields a minus sign. This can be attributed to the fact that each phonon excitation is not a direct excitation, but rather a two-photon process with an intermediate electronic excitation.

\section{Electronic degrees of freedom}
\label{electronic-DOF}
The electronic states of individual trapped ions are experimentally well controlled and can be read out with high efficiency \cite{Leibfried}. Narrow-band lasers allow the restriction of the theoretical treatment to two relevant electronic levels for each ion, separated by the carrier frequency. The collective motion of the ions can be used to create interactions between the electronic levels of different ions.
%In this section we point out how standard methods of spin manipulation and readout can be used to obtain multidimensional spectra of trapped ion spin systems.

In the following we often refer to operations on a single ion, omitting the index of the addressed ion for ease of notation. If we drive the carrier transition with the laser phase $\varphi$ for time $t$, we generate the unitary dynamics
\begin{align}\label{eq.carrierflop}
U^{\varphi}_c(t)=\cos\left(\frac{\Omega t}{2}\right)\mathbb{I}-i\sin\left(\frac{\Omega t}{2}\right)(e^{i\varphi}\sigma_++e^{-i\varphi}\sigma_-).
\end{align}
For $\Omega t=\pi/2$ (a $\pi/2$-pulse), we obtain $U^{\varphi}_c(\frac{\pi}{2\Omega}) = \frac{1}{\sqrt{2}}\mathbb{I}-\frac{i}{\sqrt{2}}\left(e^{i\varphi}\sigma_++e^{-i\varphi}\sigma_-\right)$.
%\\U_{\frac{\pi}{2}} = \frac{1}{\sqrt{2}}\mathbb{I}+\frac{1}{\sqrt{2}}\left(e^{i(\varphi-\pi/2)}\sigma_+-e^{-i(\varphi-\pi/2)}\sigma_-\right).
Introducing $\phi = \varphi-\pi/2$ yields an interaction of the form,
\begin{align}\label{eq.spinexcitation}
U^{\phi}_{\frac{\pi}{2}} = \frac{1}{\sqrt{2}}\mathbb{I}+\frac{1}{\sqrt{2}}\left(e^{i\phi}\sigma_+-e^{-i\phi}\sigma_-\right),
\end{align}
realizing the parameters $\alpha=\beta=-\gamma=1/\sqrt{2}$ and $A=\sigma_+$ in Eq.~(\ref{eq.generalint}).

A $\pi/2$ carrier pulse may thus be used to induce excitations for multidimensional spectra and the phase $\phi$ can be controlled experimentally to select the desired pathways by phase-cycling. Note that the present theoretical treatment, as can be seen from Eq.~(\ref{rho^n}), contains the impulsive limit, which only applies, if the excitation is faster than the system dynamics. For many simulated systems one can adjust the time scale of the system evolution by tuning the effective interaction strength between spins. When it is not possible or desirable to apply strong carrier pulses on much faster timescales than the system dynamics it is still possible to apply our formalism by using shorter pulses. This will reduce the amplitude for an excitation [represented by the terms proportional to $\sin(\Omega t/2)$ in Eq.~(\ref{eq.carrierflop})] and therefore reduces the signal-to-noise ratio. Since only the phase-dependent part of the signal is selected by phase cycling the amplitudes do not influence the multidimensional spectrum. However, choosing the $\pi/2$-time maximizes the signal-to-noise ratio.

We finally note that the readout of electronic populations can be implemented with nearly unit efficiency by collecting the fluorescence light while coupling one of the qubit levels to a short-lived excited state \cite{Leibfried}. This can be done for each of the ions individually, either by focussed lasers and a photo-multiplier tube or a CCD (charge-coupled device) camera. In this article we only make use of $O_j = \sigma_z^{(j)}$, where $\sigma_z^{(j)}$ denotes the Pauli $z$ matrix at spin $j$. In general, combinations of the readout method with single-qubit rotations enable to probe arbitrary observables.

\section{Vibrational degrees of freedom}
\label{vibrational-DOF}
Due to strong Coulomb interactions, the motion of different ions in a common harmonic potential is coupled and may be described using collective vibrational modes by the Hamiltonian
\begin{align}\label{eq.phononsnoU}
H=\sum_{i=1}^N\omega^0_ia^{\dagger}_ia_i+\sum_{\substack{i,j=1\\(i<j)}}^Nt_{ij}(a^{\dagger}_ia_j+a^{\dagger}_ja_i),
\end{align}
where $a^{\dagger}_i$ creates a local phonon at site $i$, $a^{\dagger}_i|0\rangle = \vert 1_i \rangle$. The order of magnitude of the average inter-ion distance is given by the length scale $l_0^3=e^2/(m\nu_z^2)$, with the axial/longitudinal trap frequency $\nu_z$ and the ion mass $m$ \cite{James}. The Hamiltonian~(\ref{eq.phononsnoU}) is a valid approximation if the parameter $\beta_0:=e^2/(l_0^3m\nu_x^2)=\nu_z^2/\nu_x^2 \ll 1$, where the radial/transverse trap frequencies are comparable, $\nu_x\approx\nu_y$, i.e., when considering a linear trap architecture \cite{PorrasCirac}. The local trap frequencies and the coupling matrix can be microscopically derived as \cite{PorrasCirac}
\begin{align} \label{eq.phonon-site-energy}
\omega^0_{i}/\nu_x&=1-\frac{\beta_0}{2}\sum_{j\neq i}\frac{1}{|u^0_i-u^0_{j}|^3},\notag\\
t_{ij}/\nu_x&=\frac{\beta_0}{2}\frac{1}{|u^0_i-u^0_{j}|^3},
\end{align}
where $u^0_i=z^0_i/l_0$ and $z_i^0$ denote the ion's equilibrium positions \cite{James}. This leads to a normalized Hamiltonian which is fully determined by the parameters $\nu_x$ and $\beta_0=\nu_x^2/\nu_z^2$. Anharmonic corrections to the potential can be induced with tunable strength $U$ and lead to an additional term $U\sum_ia^{\dagger 2}_ia_i^2$ \cite{PorrasCirac}. This results in the Bose-Hubbard Hamiltonian with long-range couplings \cite{PorrasCirac}:
\begin{align}\label{eq.phonons}
H_{\mathrm{ph}}=H+U\sum_{i=1}^Na^{\dagger 2}_ia_i^2.
\end{align}
Note that even though it is possible to locally excite or destroy phonons (as we will show in the following), resonances in nonlinear measurements will always reveal the eigenstates of the full Hamiltonian (\ref{eq.phonons}). In the spirit of nonlinear spectroscopy, we will denote the eigenstates of the single-excitation subspace as single excitons $e$,
\begin{align}\label{eq.es}
\vert e_i \rangle &= \sum_{j = 1}^N c_{i j} a^{\dagger}_j \vert 0 \rangle,
\end{align}
and the doubly-excited states by $f$,
\begin{align}\label{eq.fs}
\vert f_i \rangle &= \sum_{j,k} d_{i jk} a^{\dagger}_j a^{\dagger}_k \vert 0 \rangle.
\end{align}

\subsection{Readout}
The motional degree of freedom can be probed by analyzing the electronic spectrum. For example, if an electronic resonance of a single trapped ion is found at $\nu_0$ (carrier transition), one can usually also find a blue (red) sideband (sometimes referred to as Stokes- and anti-Stokes-line) at $\nu_0+\nu_x$ ($\nu_0-\nu_x$). If the distribution of the motional state can be assumed to be thermal, the populations are directly read out by comparison of the intensities of red and blue sidebands \cite{Leibfried}. In the present study, we consider non-equilibrium phonon distributions, and therefore need to use another readout mechanism. For states very close to the ground state we make use of an alternative scheme (see also \cite{Heinzen}). The red sideband vanishes for $T=0$. We therefore probe low-temperature excitations by driving the red sideband for a fixed time, corresponding to the length of a $\pi$-pulse on the blue sideband for the motional ground state. A blue sideband $\pi$-pulse on the ground state is reached for $t=\pi/\Omega_0^1$, where $\Omega_n^1=\eta\sqrt{n+1}\Omega+\mathcal{O}(\eta^2)$, $\Omega$ denotes the carrier Rabi frequency and $\eta$ is the Lamb-Dicke parameter \cite{Leibfried}. Application of a red sideband pulse of equal intensity and duration to an arbitrary motional Fock state $|n\rangle$ yields (we assume the electronic state to be the ground state $|\downarrow\rangle$)
\begin{align}
U_{-1}(\pi)|\downarrow,n\rangle=\cos\left(\frac{\Omega^1_{n-1}}{\Omega^1_0}\frac{\pi}{2}\right)|\downarrow,n\rangle+\sin\left(\frac{\Omega^1_{n-1}}{\Omega^1_0}\frac{\pi}{2}\right)|\uparrow,n-1\rangle.
\end{align}
We have
\begin{align}
\frac{\Omega^1_{n-1}}{\Omega^1_0}=\sqrt{n}+\mathcal{O}(\eta^2).
\end{align}
Measuring the population by fluorescence detection of the $|\uparrow\rangle$ state after application of this pulse to an arbitrary state of the form
\begin{align}
\rho=\sum_{n,m=0}^{\infty}\rho_{nm}|\downarrow,n\rangle\langle \downarrow,m|
\end{align}
is therefore equivalent to measuring the motional observable \cite{Letter}
\begin{align}
O=\sum_{n=0}^{\infty}\sin^2\left(\sqrt{n}\frac{\pi}{2}\right)|n\rangle\langle n|.
\end{align}

We assume that the ion chain has been prepared close to its vibrational ground state. Single ion addressing allows us to probe or excite (see following section) local phonons at different sites if the transitions can be carried out with sufficient laser power \cite{Brown, Harlander}. More precisely, this requires $\eta\Omega\gg \beta_0\nu_x$. The phonon hopping rate $\beta_0$ can be adjusted by variation of the ratio of axial and radial trap frequencies. %Since a large inter-ion distance poses difficulties for the implementation of sideband cooling, it may be advantageous to reduce the axial trap frequency adiabatically after the chain has been cooled to the ground state \cite{Haze}.

\subsection{Controlled excitation of local phonons}
\label{sec.phonons}
Various experimental approaches can be used to create or annihilate local phonons. Here we discuss a stimulated Raman scheme and resonant pulses on sidebands. All approaches require individual addressing of single ions in the chain. This can be achieved, for instance, by tightly focussed lasers \cite{Schindler}, by a magnetic field gradient leading to spatially dependent resonance transitions generated by Zeeman shifts \cite{Flo}, or by using different isotopes of an ion species \cite{Schaetz}.
% two possibilities to create an interaction of the form
%\begin{align}\label{eq.excitationoperator}
%D(\alpha e^{i\phi})\approx \mathbb{I}+\alpha e^{i\phi} a_i^{\dagger}-\alpha e^{-i\phi} a_i.
%\end{align}
%This interaction will create or destroy an excitation from the system with some probability $\alpha$. In this case, the phase of the field is imprinted on the quantum state. If no interaction takes place (corresponding to the identity $\mathbb{I}$) no phase shift is aquired. The possibility that no excitation is induced is important for the investigation of coherences, where only one side of the density matrix experiences an excitation while the other is subject to the identity. An important requirement for the selection of such coherences is the experimental control of the phase of the interaction field.

\begin{figure}
 \centering
 \includegraphics[width=0.49\textwidth]{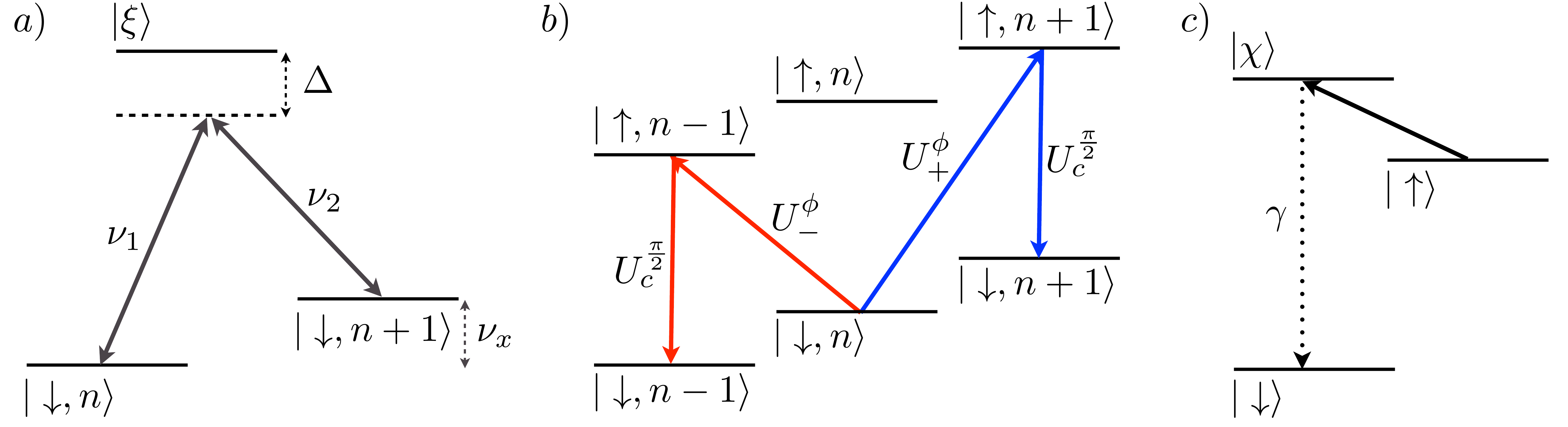}
 \caption{(Color online) a) Level scheme for the generation of a motional displacement by stimulated Raman scattering. b) Phonons can also be selectively created and destroyed by combinations of resonant sideband and carrier pulses. c) Undesired excited state population can be incoherently pumped out into the ground state via another short-lived excited state.}
 \label{fig.excitation-scheme}
\end{figure}

\subsubsection{Weak excitation by off-resonant stimulated Raman transitions}
The stimulated Raman scattering technique is routinely implemented in state of the art ion trap experiments, especially when working with a qubit encoded into two hyperfine levels of the ground state \cite{Leibfried}. These experiments bare close analogy to multidimensional spectroscopy of molecular vibrations \cite{HammZanni}. Stimulated Raman scattering off a short-lived electronic excited state $|\xi\rangle$ involves two laser fields with central frequencies $\nu_a$ and $\nu_b$ whose difference is resonant with a local vibrational transition [see Fig.~\ref{fig.excitation-scheme} a)], 
%CHANGE FIGURE ACCORDINGLY (MAYBE ADD DIAGRAMS FOR OTHER EXCITATION SCHEMES?)
\begin{align}
\nu_a - \nu_b &= \nu_{x}.
\end{align}
The two laser pulses are kept off-resonant from the electronic transition, i.e., the detuning $\Delta$ is large, such that excitations of the state $|\xi\rangle$ can be safely neglected.
%the excited state $\xi$ can be eliminated from the theoretical description, and the scattering process be described by the polarizability
%\begin{align}
%\alpha_{i, 1g} &= - \hbar \frac{\Omega_{g \xi} \Omega_{1 \xi}}{\Delta} e^{i \phi},
%\end{align}
%where $\Delta$ denotes the detuning of the two laser fields from the electronic transition. The two Rabi frequencies are given by the dipole moments connecting the two states, and the spectral envelope of the laser pulses. The phase $\phi$ consists of the phase difference between the two fields, and can be controlled by varying one of these. The polarizability depends parametrically on the central time of the laser pulses as well. If these phases are strong enough, the pulses can be chosen very short compared to typical timescales of the vibrational dynamics.\\
%{\em Alternatively:} 
%by the polarizability for the transition between the vibrational states $\vert g, n \rangle$ and $\vert g, n+1 \rangle$ of ion i is given by
%\begin{align}
%\alpha_{i} &= \sum_e \frac{\mu_{\xi n+1} \mu_{n \xi}}{\Delta} f (\omega_{vib})  e^{i \phi},
%\end{align}
% where we defined the two-photon spectral density
%\begin{align}
%f (\omega_{vib}) &= \frac{1}{2 \pi} \int  d\omega \; \epsilon_2 (\omega) \epsilon_1^{\ast} (\omega + \omega_{vib})
%\end{align}
%of the two laser envelopes $\epsilon_1$ and $\epsilon_2$. 
In this situation, the excited state $|\xi\rangle$ can be adiabatically eliminated from the theoretical description \cite{Heinzen}, and the scattering process is described by the effective coupling strength \cite{Leibfried}
\begin{align}
(\hbar/2)\Omega_{\text{eff}}=-\hbar\frac{\Omega_{n,\xi}\Omega_{n+1,\xi}}{\Delta}e^{i\phi},
\end{align}
where $\Omega_{n,\xi}$ and $\Omega_{n+1,\xi}$ denote the coupling strength of the two ground states to the excited state, and $\phi$ describes the effective phase shift induced by interaction with the two fields, which can be controlled by varying one of their phases. In nonlinear molecular spectroscopy, this quantity is refered to as the ion polarizability \cite{Shaul_book}. %The polarizability depends parametrically on the central time of the laser pulses as well. If these phases are strong enough, the pulses can be chosen very short compared to typical timescales of the vibrational dynamics.
When an ion at site $j$ is subjected to this Raman process at a timescale which is short enough to address local phonon modes, we can formally describe the action on the motional quantum state by a displacement operator \cite{Monroe96}
\begin{align}
D_j(\epsilon e^{i\phi_j})=e^{\epsilon e^{i\phi_j} a_j^{\dagger}-\epsilon e^{-i\phi_j} a_j},
\end{align}
where the displacement is given by $\epsilon=\eta\Omega_{\text{eff}}t$ with the duration $t$ of the interaction, and $a_j$ is the annihilation operator of a phonon at ion $j$. Thus, by choosing short pulses we can achieve small displacements which allow us to approximate the operator by its leading terms up to linear order:
\begin{align}\label{eq.displace}
D_j(\epsilon e^{i\phi_j})= \mathbb{I}^{(j)}+\epsilon e^{i\phi_j} a_j^{\dagger}-\epsilon e^{-i\phi_j} a_j+\mathcal{O}(\epsilon^2).
\end{align}
This operator represents one possibility to achieve an interaction operator, as shown in Eq.~(\ref{eq.generalint}). Here we have $\alpha\approx1$ and $\beta=-\gamma=\epsilon\ll1$.

\subsubsection{Strong excitation by resonant sideband pulses}
\label{sec.selectiveexcitation}
Another possibility for the creation of local phonons involves the combination of strong resonant pulses on carrier and sideband transitions. In particular, a local phonon can be created by a blue sideband pulse. Here we propose a controlled generation of phase-coherent motional excitations or de-excitations in three steps:

(i) Assuming that initially the system is prepared in the electronic ground state $|\downarrow\rangle$, a pulse on the first blue sideband for time $t$ leads to a Rabi oscillation described by
\begin{align}
U^{\phi}_+(t_1)|\downarrow,n\rangle=\cos\left(\frac{\Omega_nt_1}{2}\right)|\downarrow,n\rangle+e^{i\phi}\sin\left(\frac{\Omega_nt_1}{2}\right)|\uparrow,n+1\rangle,
\end{align}
where $\phi$ is the phase of the sideband pulse [see Fig.~\ref{fig.excitation-scheme} b)].

(ii) Next, we have to reset the electronic state back to the ground state to get the system ready for the next excitation. To do this, we employ a pulse on the carrier transition, described by $U^{\frac{\pi}{2}}_c(t_2)$ [cf. Eq.~(\ref{eq.carrierflop}) and Fig.~\ref{fig.excitation-scheme} b)], generating the state
\begin{align}
U_c^{\frac{\pi}{2}}(t_2)U^{\phi}_+(t_1)|\downarrow,n\rangle=\:&\cos\left(\frac{\Omega_nt_1}{2}\right)\cos\left(\frac{\Omega t_2}{2}\right)|\downarrow,n\rangle\notag\\
&+e^{i\phi}\sin\left(\frac{\Omega_nt_1}{2}\right)\cos\left(\frac{\Omega t_2}{2}\right)|\uparrow,n+1\rangle\notag\\
&+\cos\left(\frac{\Omega_nt_1}{2}\right)\sin\left(\frac{\Omega t_2}{2}\right)|\uparrow,n\rangle\notag\\
&+e^{i\phi}\sin\left(\frac{\Omega_nt_1}{2}\right)\sin\left(\frac{\Omega t_2}{2}\right)|\downarrow,n+1\rangle.
\end{align}
The remaining population in the $|\uparrow,n+1\rangle$ state still carries the phase shift $e^{i\phi}$ and therefore can lead to problems for subsequent excitations: A blue sideband pulse on this state, for instance, would \textit{annihilate} a motional excitation rather than \textit{create} one, and vice-versa for a red sideband pulse. We can avoid such unintended results by destroying the phase coherence of all remaining contributions of $|\uparrow\rangle$ states by pumping them back into the ground state, which is done in the third and final step of this sequence.

(iii) To pump the remaining $|\uparrow\rangle$ population incoherently back into the ground state, we use a laser to address a transition from the $|\uparrow\rangle$ state to another short-lived excited state $|\chi\rangle$ which decays rapidly back into the ground state $|\downarrow\rangle$, as depicted in Fig.~\ref{fig.excitation-scheme} c). This finally leads to the density matrix
\begin{align}\label{eq.rhofinal}
\rho_f=\:&|\downarrow\rangle\langle\downarrow|\otimes
\left[\cos^2\left(\frac{\Omega_nt_1}{2}\right)|n\rangle\langle n|\right.\notag\\
&\hspace{1.3cm}+\frac{1}{2}e^{-i\phi}\sin(\Omega_nt_1)\sin(\Omega t_2)|n\rangle\langle n+1|\notag\\
&\hspace{1.3cm}+\frac{1}{2}e^{i\phi}\sin(\Omega_nt_1)\sin(\Omega t_2)|n+1\rangle\langle n|\notag\\
&\left.\hspace{1.3cm}+\sin^2\left(\frac{\Omega_nt_1}{2}\right)|n+1\rangle\langle n+1|\right].
\end{align}
We must briefly discuss the recoil energy, which may induce unwanted changes to the ion's motional state. During the repump process one photon is absorbed and another emitted, leading to a maximal increase in kinetic energy of
\begin{align}
E_{\text{recoil}}=\frac{\hbar^2(k_1^2+k_2^2)}{2m},
\end{align}
while the potential energy of a low-$n$ Fock state $|n\rangle$ of a harmonic oscillator is of the order of
\begin{align}
E_{\text{h.o.}}=\hbar\nu_x.
\end{align}
Using experimental parameters for $^{40}\text{Ca}^+$ ($k_i=2\pi/\lambda_i$ with $\lambda_1=854$\,nm and $\lambda_2=397$\,nm), we find that $E_{\text{recoil}}/\hbar\approx 242$\,kHz while typically $\nu_x\approx 2\pi\times3$\,MHz. The recoil has therefore negligible effect on the motional state.

The final state $\rho_f$ has all the required properties to be useful for a nonlinear measurement protocol [see also Eq.~(\ref{eq.interact})]: The coherences have the correct phase dependence for selection by phase cycling and contain all excitations created by a single interaction, while the de-excitations do not occur. Furthermore, the electronic state is fully reset to the ground state, making it susceptible for another identical excitation at a later time. For the selection of individual pathways the duration of the two pulses is not particularly important, since only the contribution with the appropriate phase-shift will be selected later to obtain the desired diagram. However, in order to maximize the utilizable signal, generated by the phase-dependent motional coherences, it is desirable to optimize the pulse lengths. From Eq.~(\ref{eq.rhofinal}) we see that the signal yield is maximized when both $\sin(\Omega_nt_1)=\sin(\Omega t_2)=1$, corresponding to $\pi/2$-pulses with durations $t_1=\pi/2\Omega_n$ and $t_2=\pi/2\Omega$, respectively. Since the $\pi/2$-time of the sideband pulse $U_+^{\phi}(t_1)$ depends on $n$ and in principle would have to be obtained by measurement each time, for simplicity it may be advantageous to set the duration to a fixed value, e.g., $t_1=\pi/2\Omega_0$. In the applications discussed in the present work we focus on low-temperature motional states, i.e., the most relevant contribution is generated by the ground state $|0\rangle$ or Fock states close to the ground state, allowing us to fix the pulse duration as described before.

Since the decay in step~(iii) happens very fast (typically on the order of $10$ nanoseconds) compared to the vibrational dynamics (microseconds), the dominant contribution to the entire three-step generation of vibrational excitations stems from the sideband ``excitation" pulse [step~(i)] and the carrier ``reset" pulse [step~(ii)]. The sideband pulse typically is one order of magnitude slower than the carrier pulse due to the small Lamb-Dicke parameter $\eta$. Thus, both pulses happen on a faster timescale than the phonon hopping if the condition
\begin{align}
\frac{2\Omega\eta}{\pi}\gg\beta_0\nu_x
\end{align}
is satisfied. In this case, we can describe the effective action on the vibrational degrees of freedom as instantaneous, leading to an interaction operator [cf. Eq.~(\ref{eq.generalint})] of the form
\begin{align}\label{eq.vplus}
V_+=\alpha\mathbb{I}+\beta e^{i\phi}a^{\dagger},
\end{align}
where $\alpha$ and $\beta$ are positive, but may depend on the initial motional occupation. Again, their precise value (other than their relative signs) does not play an important role for the signal after phase cycling.

Correspondingly, a de-excitation can be generated by replacing the blue-sideband pulse in step~(ii) with a pulse on the red sideband, $U^{-\phi}_-(t)$, which couples the pairs of states $|\downarrow,n\rangle$ and $|\uparrow,n-1\rangle$. Eventually, analogous considerations as above lead to the effective transition operator
\begin{align}\label{eq.vminus}
V_-=\alpha\mathbb{I}+\gamma e^{-i\phi}a.
\end{align}

The above methods allow for the selective excitation or de-excitation of a local phonon, that is, we can design schemes to generate interactions of the form of Eq.~(\ref{eq.generalint}) such that $\beta=0$ or $\gamma=0$. We emphasize that this is not possible in conventional spectroscopy experiments, e.g., when addressing molecular aggregates by direct dipole transitions. The available control for cold matter systems thus provides us with an important advantage, of which we make explicit use in Sec.~\ref{sec.singlediagrams}.

We can also use sideband pulses to create a displacement  operator, in analogy to the Raman process discussed in the previous section. To this end, we drive both red and blue sidebands simultaneously, similarly to the M\o lmer-S\o rensen gate \cite{MS}, but here we consider the pulses to be on resonance. This generates a displacement $D(\alpha e^{i\phi})$ of the motional states \cite{MS,Monroe10}, where we assume that both sidebands are driven with the same intensity $\eta\Omega$. We further assume that blue and red sideband have the same phase with opposite signs: $\phi_r=-\phi_s=\phi+\pi/2$. The amplitude of the displacement is then given by $\epsilon=\eta\Omega t/2$ and effective phase shift is given by $\phi$ \cite{Monroe10}.

\section{Single quantum coherence signals}

\begin{figure*}[tb]
\centering
\includegraphics[width=\textwidth]{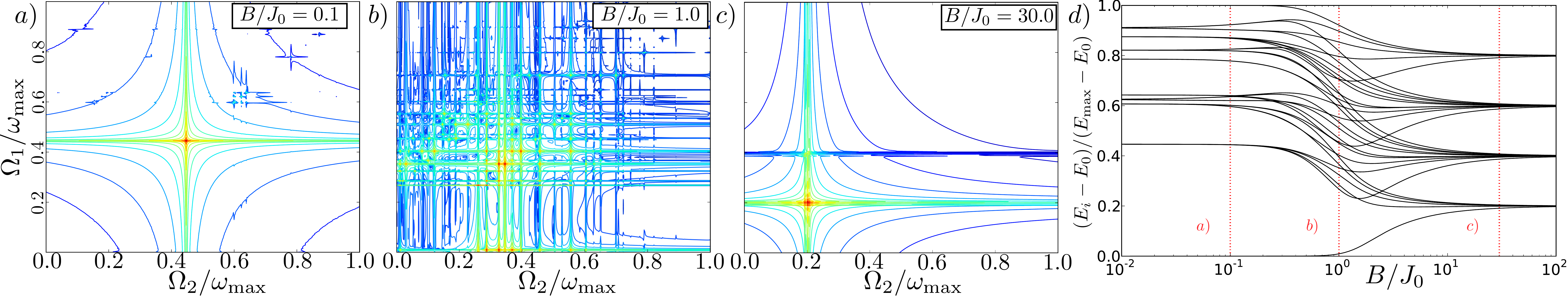}
\caption{(Color online) a-c) Positive frequency sector of the \textit{SQC} signal $\mathrm{arcsinh}\vert S^{\textit{SQC}}_{11;1} (\Omega_1, \Omega_2) \vert$ in an Ising spin chain of five ions for different values of $B/J_0$. d) The normalized spectrum shows clear signatures of the emerging quantum phase transition from ferromagnet to paramagnet for increasing $B/J_0$. A spin flip far from criticality [a) and c)] predominantly couples two neighboring state manifolds. In contrast, the quantum critical regime is characterized by lack of good quantum numbers and broadly distributed energy levels. Any perturbation generically couples to the entirety of excited states, as reflected by the broad peak spectrum in the two-dimensional spectrum b). Throughout this manuscript, the color code shows the relative strengths of individual two-dimensional peaks increasing from blue to red on a linear scale.}
\label{fig.spinqpt}
\end{figure*}

\label{sec.sqc}
As described in Ref.~\cite{Letter}, the \textit{single quantum coherence (SQC)} signal involves two pulses followed by fluorescence detection, hence, a second-order signal with two variable time delays. Using the methods developed in Sec.~\ref{theory}, we select the phase signature $\phi_1 - \phi_2$ from the entire second-order signal by phase cycling [Eq.~(\ref{eq.2ndorder})]. The resulting signal $S^{\textit{SQC}}_{i_1,i_2;j}(t_1,t_2)$ is ideally suited to reveal the excitation spectrum and coupling terms, as well as environmental influences. It can separate coherent from incoherent transport processes, and assess and quantify noise processes \cite{Letter}. Here, we demonstrate its versatility by demonstrating two further applications:
\begin{itemize}
	\item the detection of steady state currents,
	\item the observation of complex excited state dynamics near a quantum phase transition.
\end{itemize}

The different contributions to the \textit{SQC} signal are depicted in Fig.~\ref{fig.sqc}. During the first time delay $t_1$, we select the contribution of a coherence between the initial state and a single excited state [cf. also Eq.~(\ref{eq.sqcleftmost})]. When the initial state is given by the ground state of the system and only excited-state populations are measured, we can restrict to the leftmost diagram, and expanding the selected coherence in a basis of energy eigenstates $H|e_i\rangle=\omega_i|e_i\rangle$ yields
\begin{align}
A^{\dagger}_{i_1}|0\rangle\langle 0|=\sum_{j}c_{ji_1}^*|e_j\rangle\langle 0|.
\end{align}
During $t_1$ each of these eigenstates picks up a phase, 
\begin{align}
\sum_{j}c_{ji_1}^*\mathcal{G}(t_1)|e_j\rangle\langle 0|=\sum_{j}c_{ji_1}^*e^{-i\omega_jt_1}|e_j\rangle\langle 0|,
\end{align}
such that a Fourier transform of the \textit{SQC} signal,
\begin{align}\label{eq.sqcft}
S^{\textit{SQC}}_{i_1,i_2;j}(\Omega_1,\Omega_2)=\int\limits_0^{\infty}dt_1\int\limits_0^{\infty}dt_2S^{\textit{SQC}}_{i_1,i_2;j}(t_1,t_2)e^{-i\Omega_1t_1}e^{-i\Omega_2t_2},
\end{align} 
reveals the entire single-excitation spectrum along the $\Omega_1$ axis \cite{Letter}. During the second time delay $t_2$, the measured signal stems from coherent superpositions of single excited states, and Eq.~(\ref{eq.sqcft}) consequently reveals differences of single-exciton frequencies along $\Omega_2$. For different initial states, other contributions also need to be taken into account, but the steps above can be followed analogously to interpret the signal. By employing single-site addressability for local readout, we can use the double Fourier transformed \textit{SQC} signal (\ref{eq.sqcft}) to obtain information on the excitation transfer and the couplings within a chain of ions.

\subsection{Probing Excitation Spectrum and Couplings across a Quantum Phase Transition}
We now apply the \textit{SQC} signal to study the excitation spectrum and couplings in a spin chain with different values of system parameters, which may be associated with different quantum phases. We consider a chain of spins subject to long-range Ising-type interactions $J_{ij}$ and a global magnetic field $B$. The Hamiltonian 
\begin{align}
H_{\mathrm{sp}}= -\sum_{\substack{i,j=1\\(i < j)}}^N J_{ij} \sigma_x^{(i)} \sigma_x^{(j)} - B \sum_{i=1}^N \sigma_y^{(i)}
\end{align}
can be created by coupling the electronic degrees of freedom of a chain of trapped ions via their common motional modes by applying appropriate laser fields \cite{PorrasCiracSpins}. The interactions decay algebraically with the distance as $J_{ij}\approx J_0/|i-j|^{\alpha}$, with $\alpha\in(0,3)$. For $B\gg J_0$ the model describes a paramagnet, meaning that in the ground state, all spins align along the direction of the global magnetic field. When $B\ll J_0$ the system behaves (anti-)ferromagnetically when ($J_0<0$) $J_0>0$. In the crossover region, where both spin-spin couplings and global magnetic field are of comparable order of magnitude, a quantum phase transition \cite{QPTs} occurs. This phase transition is characterized by a discontinuous change of the ground state in the thermodynamic limit. Nevertheless, clear signatures thereof can be observed in finite-sized systems \cite{Islam11}.

As can be seen in Fig.~\ref{fig.spinqpt} d), the ground state is formed by a degenerate manifold of states for $B / J_0 \ll 1$, and splits up as $B / J_0$ is increased. However, dramatic changes can also be observed in the excitation spectrum, such as broadly distributed energy levels and strong couplings between different eigenstates close to the critical point, as revealed by avoided crossings \cite{Madronero}. These phenomena typically indicate competing symmetries of comparable strengths \cite{Victor}, and the onset of the resulting macroscopic signature can often be detected already in relatively small systems \cite{AbuEPL}. Such effects are quite generic and can be observed in a variety of quantum optical models, including the Bose-Hubbard model \cite{Venzl}, which we will discuss later in a different context. As we will demonstrate below, such phenomena can be probed conveniently with nonlinear spectroscopy.

In Fig.~\ref{fig.spinqpt} a-c), we display the \textit{SQC} signal of a chain of five ions with $\alpha = 1$, initially prepared in its ground state for different parameters of $B/J_0$. The induced excitations correspond to local spin flips at the left end of the spin chain, described by a $\sigma_+^{(1)}$ operator [cf. Eqs.~(\ref{eq.generalint}) and (\ref{eq.spinexcitation})], and also the readout, corresponding to measurement of $\sigma_z^{(1)}$ is carried out at the same spin. The experimental implementation of such excitations to the electronic degree of freedom of a trapped ion is described in Sec.~\ref{electronic-DOF}. Note that, here, a de-excitation corresponds to $\sigma_-^{(1)}$, which, in contrast to systems of harmonic oscillators, can lead to non-zero contributions when applied to the ground state. We therefore cannot single out the left-most diagram of Fig.~\ref{fig.sqc}, but rather have to take the sum of all four pathways, taking into account their relative signs. Considering the interaction as described in Eq.~(\ref{eq.spinexcitation}), we obtain $\beta=-\gamma=1/\sqrt{2}$. We thus have to substract the contributions of the signals $S^{(\textit{Ll})}$ and $S^{(\textit{rR})}$ from $S^{(\textit{LR})}$ and $S^{(\textit{rl})}$ [see Eq.~(\ref{eq.totalsqc})]. Readout is performed by measuring the expectation value of $\sigma^{(1)}_z$, as described in Sec.~\ref{electronic-DOF}. The time evolution between pulses is then governed by the unitary dynamics $\rho(t)=\mathcal{G}(t)\rho(0)$, with the Green's function $\mathcal{G}(t)=e^{-iH_{\mathrm{sp}}t}$.

The two-dimensional spectra in Fig.~\ref{fig.spinqpt} can be obtained by controlling only a single spin of the chain. Far away from the phase transition, i.e., for $B\ll J_0$ and $B\gg J_0$, the excitation spectrum of the local spin flip is concentrated around a narrow energy range (y-axis) and only few coupling terms (x-axis) can be observed. For example, when $B\gg J_0$, the ground state is approximately given by $|\Psi_0\rangle=|\uparrow_y\rangle^{\otimes N}$. Roughly speaking, the first band of excited states is spanned by the set of states with one out of $N$ spins pointing upwards along the $B$-field direction, while all others are pointing down. The local spin flip along the $z$-direction thus creates a superposition between the ground state and first excited states, and only small populations in the second band are created due to remaining spin-spin couplings generated by nonzero $J_0$. Small corrections due to the non-vanishing magnetic field may be observed on the upper right sector of Fig.~\ref{fig.spinqpt} a). They may be regarded as a precursor of the transition to quantum chaos near the quantum phase transition. Close to the transition excitations are spread out over the entire energy landscape and a rich collection of couplings can be observed, indicating strong correlations and critical behavior \cite{QPTs}. In this parameter range, good quantum numbers to characterize the quantum states do not exist [see Fig.~\ref{fig.spinqpt} d)], and the set of coefficients in any representation generically has a large entropy \cite{Madronero}, which is clearly reflected by the two-dimensional spectrum for the excited state, Fig.~\ref{fig.spinqpt} b).

\subsection{Spectroscopic signatures of steady-state currents}
Trapped ions hold great potential as quantum simulators \cite{Plenio13} for quantum transport theories \cite{Znidaric, Wellens}. In this section, we demonstrate how the presence of steady state currents and their decomposition into coherent and incoherent contributions can be detected -- an important step towards the certification of such simulations.

We consider an open chain of harmonic oscillators, where the outer oscillators are connected to thermal reservoirs with different temperatures. The temperature gradient along the chain competes with the coherent couplings between the sites and eventually drives the system into a steady state. Here we aim to investigate this environment-induced asymmetry and the resulting excitation transport process using nonlinear spectroscopy.

Such a system may be modeled on the basis of the vibrational degrees of freedom in a chain of trapped ions, as described in Sec.~\ref{vibrational-DOF} (cf. also Ref.~\cite{Plenio13}). For further details on the implementation of controlled dissipation on a trapped ion, see Refs.~\cite{Poyatos,Myatt}. %The energy current between ions two adjacent sites is given by the expectation value of the Hermitian operator \cite{Prosen}
%\begin{align}
%J_{ij} &=  i \left[ H_{j-1}, H_j \right],
%\end{align}
%where $H_j$ denotes the local Hamiltonian of site $j$.

\begin{figure}
\centering
\includegraphics[width=.49\textwidth]{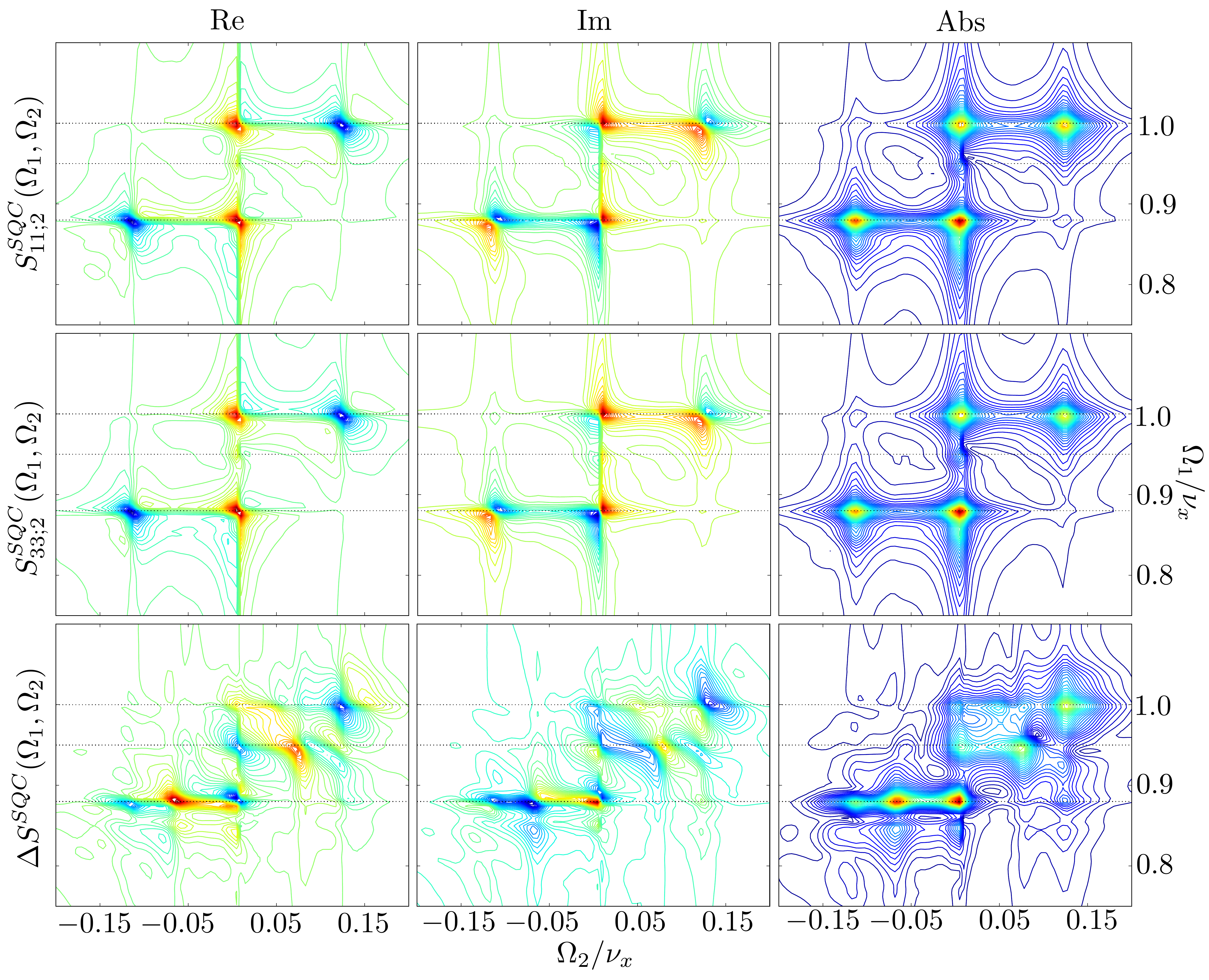}
\caption{(Color online) The \textit{SQC} signals $S^{\textit{SQC}}_{11;2}$ and $S^{\textit{SQC}}_{33;2}$ and their difference $\Delta S^{\textit{SQC}}$ for a chain of three ions subject to a temperature gradient with parameters $\overline{n}_1=0$, $\overline{n}_3=0.5$, $\gamma=0.01\nu_x$ and an anharmonic modulation $U=-0.025\nu_x$ of the trap potential. The presence of the steady state current slightly shifts the peaks in $S^{\textit{SQC}}_{11;2}$ and $S^{\textit{SQC}}_{33;2}$ (first two rows) with respect to one another, since it enhances transport along the current direction, and hinders transport opposed to it. These shifts create the signal $\Delta S^{\textit{SQC}}$.}
\label{fig.deltasqc}
\end{figure}

We apply the \textit{SQC}-protocol after the system has evolved into the steady-state. The outer ions ($i=1$ and $i=N$) are in contact with a thermal bath. The system dynamics $\rho(t)=\mathcal{G}(t)\rho(0)=e^{\mathcal{L}t}\rho(0)$ is modeled by the Lindblad master equation $\dot{\rho}(t)=\mathcal{L}\rho(t)$ as follows,
\begin{align}
\dot{\rho}(t) = -i[H_{\mathrm{ph}},\rho(t)] + \:&\gamma\sum_{i\in\{1,N\}}( \overline{n_i} + 1)\left(a_i\rho(t) a_i^{\dagger}-\frac{1}{2}\{a_i^{\dagger}a_i,\rho(t)\}\right) \notag \\
+\:& \gamma\sum_{i\in\{1,N\}} \overline{n_i} \left(a^{\dagger}_i\rho(t) a_i-\frac{1}{2}\{a_ia^{\dagger}_i,\rho(t)\}\right), \label{eq.lindblad}
\end{align}
where the Hamiltonian is given in Eq. (\ref{eq.phonons}). Here, $\gamma$ denotes the coupling strength, and $\overline{n}_i$ the mean occupation number of the local bath at ion $i$. %For our investigation, we chose $\overline{n_1} = 0$, and $\overline{n_N} = 0.5$.

To probe transport processes along the chain we apply both excitations on one end of the chain and probe the excitation at the center, measuring either $S^{\textit{SQC}}_{11;(N-1)/2}$ or $S^{\textit{SQC}}_{NN;(N-1)/2}$, when $N$ is odd. Without the temperature gradient, both signals are equal, due to the spatial symmetry of the system. Hence, the difference of the two signals
\begin{align}
\Delta S^{\textit{SQC}}=S^{\textit{SQC}}_{11;(N-1)/2}-S^{\textit{SQC}}_{NN;(N-1)/2}
\end{align}
vanishes for perfectly symmetric conditions. If, however, a temperature gradient is present, we do not expect $\Delta S^{\textit{SQC}}$ to vanish, and the signal allows us to observe preferential transport pathways and mechanisms. In Fig.~\ref{fig.deltasqc}, we show simulations of $\Delta S^{\textit{SQC}}$ for the simplest nontrivial case of $N=3$ ions, with $\overline{n}_3=0.5$, while the left-most ion is in contact with a zero-temperature bath, $\overline{n}_1=0$, and $\gamma=0.01\nu_x$. 

Since the initial state is a non-equilibrium steady-state, rather than the ground state, again, we have to take into account all four $\textit{SQC}$ diagrams shown in Fig.~\ref{fig.sqc}. Considering weak displacement operators, as in Eq.~(\ref{eq.displace}), as interaction operators, leads to $\beta=-\gamma \ll 1$ [cf. Eq.~(\ref{eq.generalint})]. Hence, we obtain the same relative signs for the total \textit{SQC} signal, Eq.~(\ref{eq.totalsqc}), as in the previous section.

For the analysis of the \textit{SQC}-signals, shown in Fig.~\ref{fig.deltasqc}, it is instructive to first consider the normalized Hamiltonian (\ref{eq.phonons}), whose single exciton sector reads for three ions
\begin{align}
H &= \begin{pmatrix}
1 - t_{12} - t_{13} & t_{12} & t_{13} \\
t_{12} & 1 - 2 t_{12} & t_{12} \\
t_{13} & t_{12} & 1 - t_{12} - t_{13}
\end{pmatrix}.
\end{align}
Diagonalization yields the single exciton energies $e_1 = 1 - 3 t_{12}$, $e_2 = 1 - t_{12} - 2 t_{13}$, and $e_3 = 1$. Hence, we note that the exciton differences $\omega_{ij}=e_i-e_j$, which in the \textit{SQC} signal can be observed along the $\Omega_2$-axis, are given by combinations of the two coupling terms $t_{12}$ and $t_{13}$. These values can be obtained, for instance, from the absolute value of $S^{\textit{SQC}}_{11; 2}$ (top row of Fig.~\ref{fig.deltasqc}). Additional strong resonances along $\Omega_2 = 0$ can be attributed to incoherent, environment-induced transport \cite{Letter}. 

\begin{figure}[tb]
\centering
\includegraphics[width=.49\textwidth]{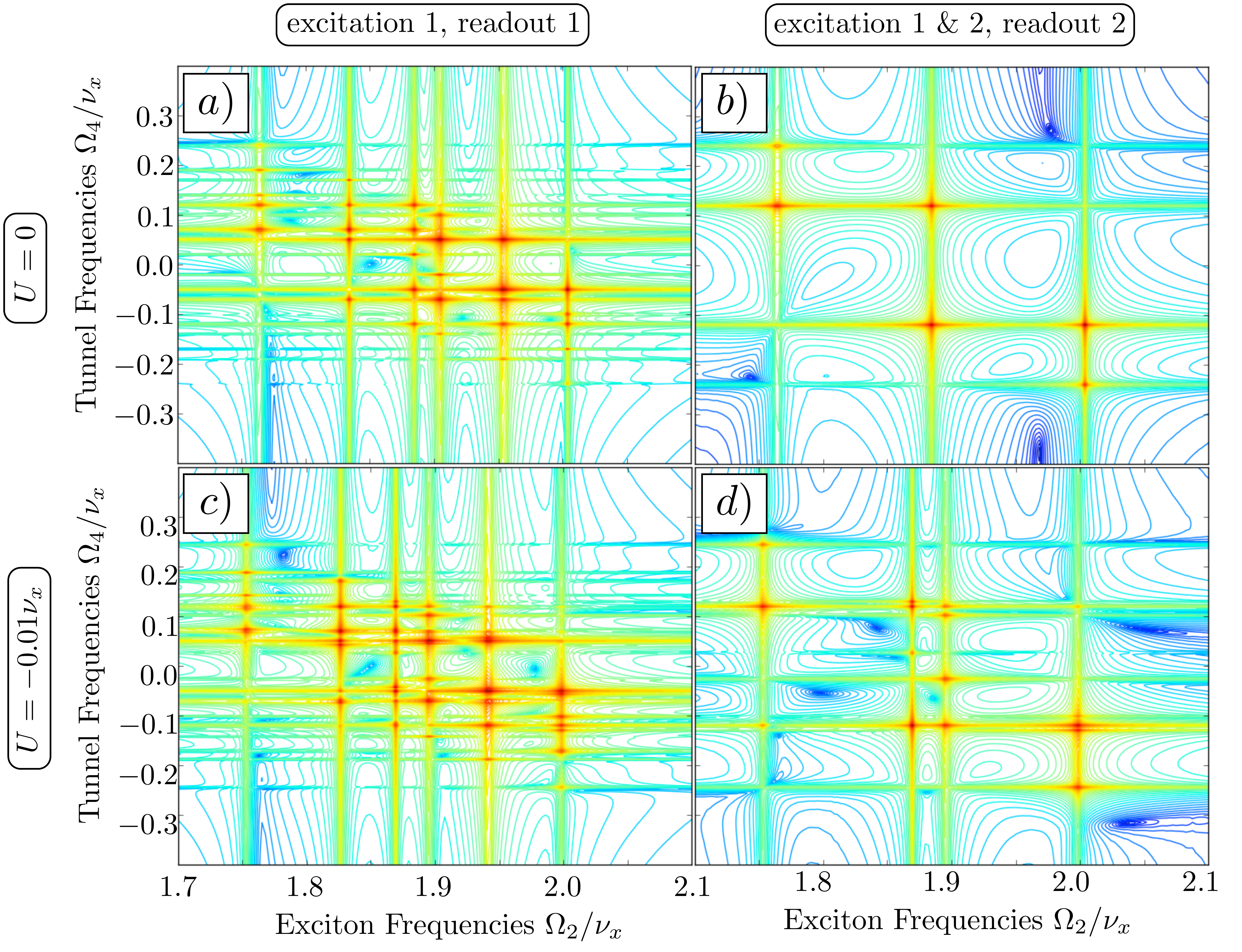}
\caption{(Color online) The \textit{DQC} signal $\mathrm{arcsinh}\vert S^{\textit{DQC}}_{i_1 i_2 i_3 i_4; j} (t_1 = 0, \Omega_2, t_3 = 0, \Omega_4) \vert$ for $U$ = 0 (top row), and $-0.01 \nu_x$ (bottom row) in a chain of three ions. The excitation and readout ions are $i_1 = i_2 = 1$ for all ions, and  either $i_3 = i_4 = j = 1$ in the left column, or $i_3 = i_4 = j = 2$ in the right column. This selection allows to ``filter" the signal, and obtain spectra of small subsets of excited states which are easier to interpret.}
\label{fig.3}
\end{figure}

We now turn to $\Delta S^{\textit{SQC}}$ in the bottom row. The first observation made for the individual $\textit{SQC}$-signals is the strong contribution of peaks at $\Omega_2\neq 0$, indicating the important role of coherent processes for excitation transport in this system. This can be explained by the relatively weak coupling to the thermal fields in comparison to the coherent intra-chain couplings. However, the presence of peaks at $\Omega_2=0$ tells us that incoherent processes also contribute significantly to the dynamics. From the difference $\Delta S^{\textit{SQC}}$ we find that most of the asymmetry is generated by coherent terms, whereas the incoherent peaks almost cancel each other entirely.

The absolute difference between the two \textit{SQC}-signals for the chosen set of parameters is of the order of 5\% of the highest peak in each individual signal \cite{uncertaintyfootnote}. Application of stronger temperature gradients, however, can reduce the required fidelity by producing stronger contrasts.

By comparing relative intensities of the two-dimensional peaks, we find that some coupling frequencies are more strongly suppressed than others, which can be attributed to a preferred direction for excitation transport, namely when following the temperature current. More precisely, the temperature current effectively leads to faster transport in the one case, and to slower transport in the other, which translates into slightly shifted tunnel frequencies, which are revealed along the $\Omega_2$-axis. The difference of two shifted dispersive peaks, as can be observed most prominently in the imaginary part of the spectrum, leads to an absorptive peak at the spot of maximal deviation. Most pronounced peaks of this form can be seen in $\Delta S^{\textit{SQC}}$ at $(\Omega_1,\Omega_2)=(\omega_3,\omega_{32})$ and $(\omega_2,\omega_{23})$, where the real part of the latter is enhanced at the expense of the former, indicating faster transport along the temperature gradient.

\begin{figure}
\centering
\includegraphics[width=.49\textwidth]{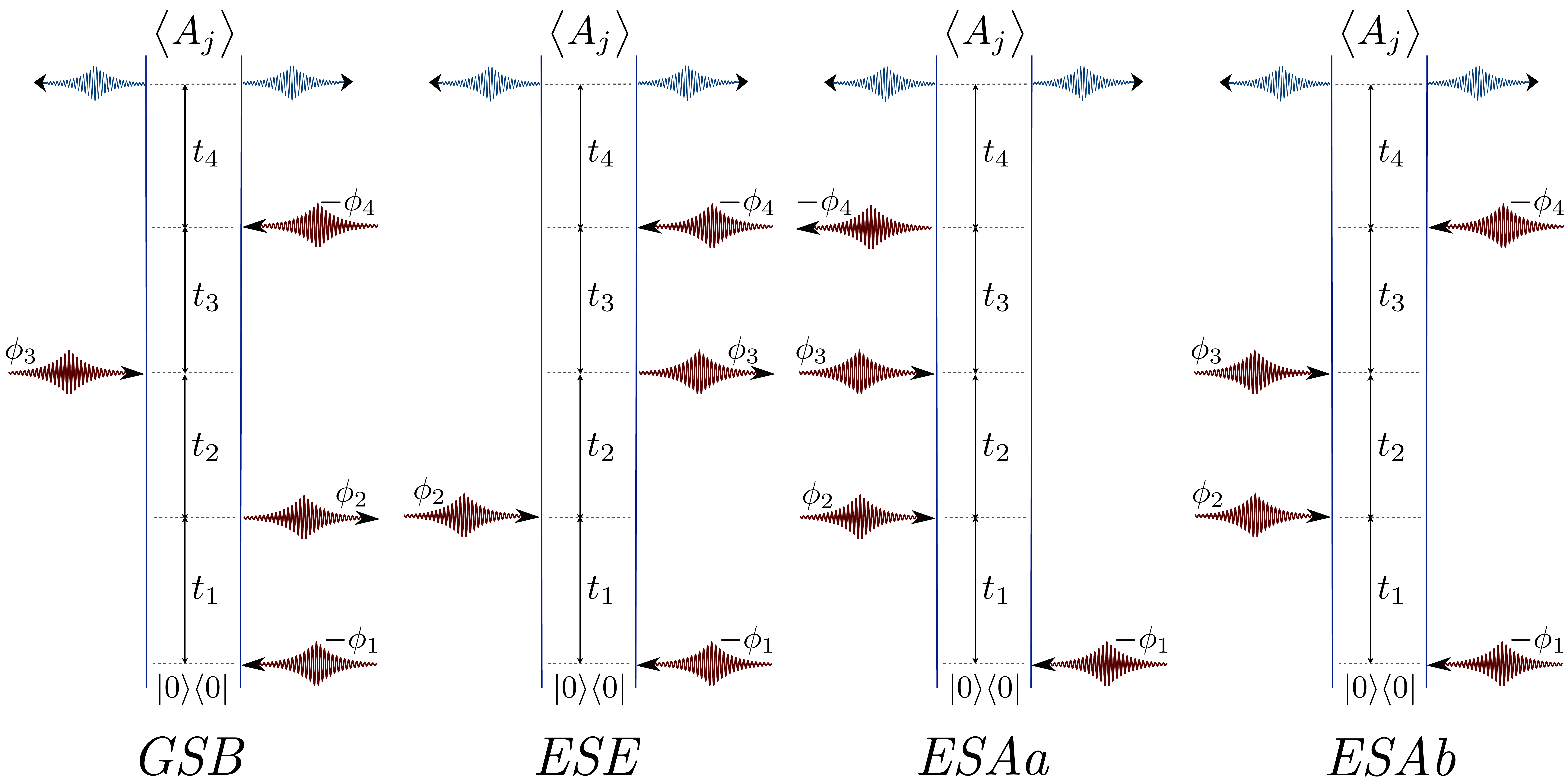}
\caption{(Color online) The diagrams with phase signature $- \phi_1 + \phi_2 + \phi_3 - \phi_4$ comprising the \textit{photon echo}. For historical reasons, they are denoted \textit{ground state bleaching (GSB)}, \textit{excited state emission (ESE)}, and \textit{excited state absorption (ESAa} and \textit{ESAb}).}
\label{fig.photon-echo-diagram}
\end{figure}

\begin{figure*}
\centering
\includegraphics[width=\textwidth]{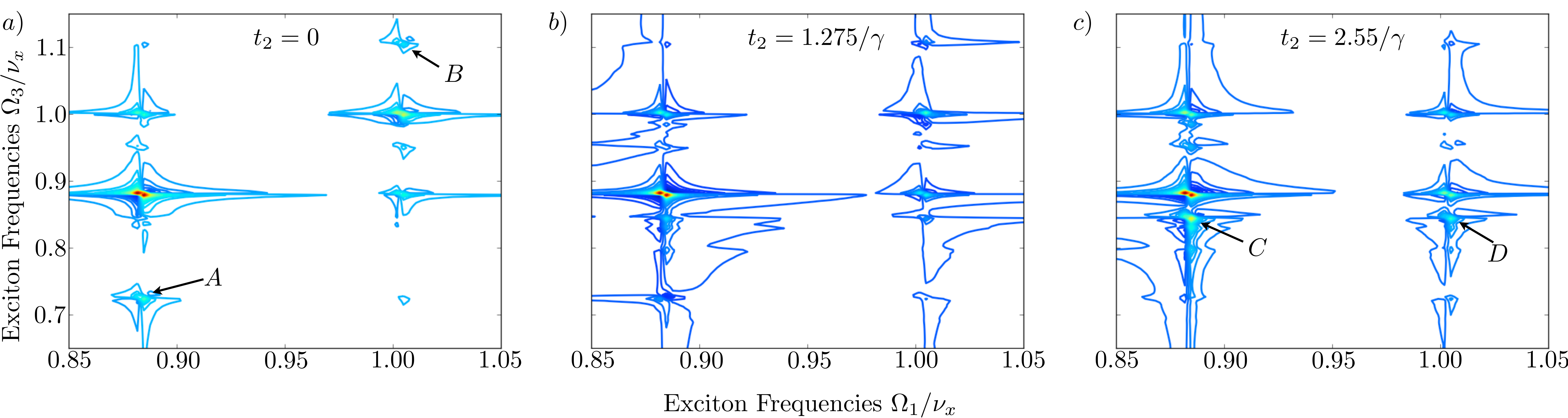}
\caption{(Color online) The real part of the \textit{photon echo} signal $ S^{\textit{PE}}_{1111; 1} (\Omega_1, t_2, \Omega_3, t_4 = 0)$ for a thermal bath with temperature $\overline{n} = 0.1$, anharmonicity $U = -0.03 \nu_x$, and population time a) $t_2 = 0$, b) $t_2 = 1.275/ \gamma$, and c) $t_2 = 2.55 / \gamma$. The anharmonic peaks $A$ and $B$ lose intensity with increasing population time, and new peaks $C$ and $D$ gain strength. Both effects can be related to environment-induced population decay, which affects the relative strength of the contributing Feynman diagrams.}
\label{fig.photon-echo}
\end{figure*}

%\begin{figure*}
%\centering
%\includegraphics[width=\textwidth]{global-thermal-bath.pdf}
%\caption{The \textit{ESAb}-diagram of the photon echo signal  $\Im \{ S^{\textit{ESAb}}_{2222; 2} (\Omega_1,t_2,\Omega_3, t_4 = 0) \}$ for a global thermal bath with temperature $\overline{n} = 0.1$, anharmonicity $U = -0.03 \nu_x$, and population times (a) $t_2 = 0$, (b) $t_2 = 385 / \nu_x$, and (c) $t_2 = 380 / \nu_x$.}
%\label{fig.global-bath}
%\end{figure*}

\section{Double quantum coherence}
\label{sec.dqc}
To explore dynamics involving multiple excitations, we now consider the \textit{double quantum coherence (DQC)} signal as defined in Ref.~\cite{Letter}.

%\subsection{Multi-exciton transport}
%\label{sec.multiexciton}
%As described before, we measure all quantities in units of the radial trap frequency $\nu_x$. 
In the following we present simulations of signals of the vibrational degrees of freedom for a chain of three ions trapped in a potential with $\beta = 0.1$. 
According to Eqs. (\ref{eq.phonon-site-energy}) and (\ref{eq.phonons}), we then obtain the single exciton energies $\omega_1 = 0.88 \nu_x$, $\omega_2 = 0.95 \nu_x$, and $\omega_3 = \nu_x$.
%\begin{table}[h!!!]
%  \centering
%  \begin{tabular}{c|ccc}
%    State  &$e_1$ & $e_2$ & $e_3$ \\ \hline
%    Energy $/ \nu_x$ & 0.88 & 0.95 & 1 \\ 
%  \end{tabular}
%  \caption{Numerical values of the single exciton states for a 3-ion chain with the ratio $\beta = 0.1$.}
%  \label{tab:three-excitons}
%\end{table}
Making use of the terminology for single excited states introduced in Eq.~(\ref{eq.es}), we identify $e_3$ with the center-of-mass mode, and $e_2$ with the breathing mode \cite{James}. These values can be used to assess the character of the multi-exciton states that appear in the following. For instance, if a two-exciton state is close to the sum of $e_2 + e_3$, one can conclude that for small anharmonicity $U$, its properties can be derived from those of the center-of-mass and the breathing mode. 

As described in Ref.~\cite{Letter}, the \textit{DQC} signal consists of a four-pulse measurement with phase signature $\phi_1 + \phi_2 - \phi_3 - \phi_4$, and time delays $t_1, \dots, t_4$. It can be used to detect anharmonicities in the trap potential, and observe couplings induced by this anharmonicity. In contrast to Ref.~\cite{Letter}, we here set the time delays $t_1$ and $t_3$ to zero, and Fourier transform the delays $t_2$ and $t_4$. Due to their positive phase shifts, the first two pulses both create excitations which only yield signals when they excite on the \textit{ket} side (see Sec.~\ref{theory}). Hence, the Fourier transform of $t_2$, i.e., $\Omega_2$, yields coherences between two-exciton states and the ground state, $f - g$, where $f$ represents some double-excited state as introduced in Eq.~(\ref{eq.fs}) and $g$ the ground state. During $t_4$, the diagrams evolve either in $f - f'$ or $e - e'$ coherences, and their Fourier transform therefore reveals tunnel rates between the different ions.
Fig.~\ref{fig.3} compares the absolute value of this signal for harmonic and anharmonic potentials (top and bottom row, respectively), and for different readout ions (left and right column). In panel a), all the excitations and the readout are carried out on the leftmost ion, such that all six possible two-exciton frequencies are resolved along $\Omega_2$, as well as all the possible tunnel rates along $\Omega_4$. Adding anharmonicity to the system [panel c)] provides an even richer peak pattern, as it leads to an interaction-induced lifting of degeneracies, which effectively splits up some of the tunnel rates. In principle, each peak can be assigned to certain excitations and corresponding ion-ion couplings. The vast number of frequencies present already in this rather small system, however, renders this a rather laborious task.

Nevertheless, the possibility to locally excite and read out phonons can be used to disentangle the spectra, and resolve small changes due to anharmonicities. Panels b) and d) depict the signals where the first two interactions excite the leftmost ions, and the following two pulses as well as the readout pulse interact with the center ion. For a harmonic potential, only those two-exciton states that do not contain the breathing mode yield a signal, due to the vanishing overlap of the breathing mode with the center ion. While Fig.~\ref{fig.3} a) shows resonances for all six $f$-states along $\Omega_2$, only three of them ($f_1, f_3$ and $f_6$) are left in Fig.~\ref{fig.3} b). Furthermore, only resonances of the $e$-states $e_1$ and $e_3$ can be observed due to the readout.
Thus, only six well-separated peaks in total constitute the signal in panel b). When we add anharmonicity in panel d), we can monitor the splitting of the degenerate resonances at $\Omega_2 = 2 \nu_x$ and $\Omega_4 = - 0.1 \nu_x$ into three closely neighboring lines. For $U = 0$, the coherence between $e_3 - e_1$ coincides with the $f$-coherences $f_6 - f_3$, $f_5 - f_2$, and $f_3 - f_1$ which are all suppressed due to the states' symmetry. As explained in Ref.~\cite{Letter}, the state $f_4$, which has no overlap with the center ion, picks up contributions from the other unperturbed two-exciton states, and therefore becomes bright in this measurement. This can also be seen more obviously in the appearance of a fourth line of resonances at $\Omega_2 \approx 1.9 \nu_x$.

\section{Detecting population decay in the photon echo}
\label{sec.photon-echo}
We finally employ the \textit{photon echo (PE)} signal to monitor the decay of excitations to a heat bath. The \textit{PE} is the most prominent four-wave mixing signal in multidimensional optical spectroscopy, where it is routinely used to monitor exciton and charge transport in aggregates \cite{HammZanni, Shaul_book, Brixner}. In our theory, it corresponds to the four-pulse measurement with phase signature $- \phi_1 + \phi_2 + \phi_3 - \phi_4$ (see Fig.~\ref{fig.photon-echo-diagram}). The so-called ``coherence times" $t_1$ and $t_3$ are Fourier transformed, and the ``population time" $t_2$ is varied to monitor transport processes. The final delay $t_4$ is set to zero. Here, we will use the \textit{PE} to monitor excitation decay into an environment.

In Fig.~\ref{fig.photon-echo} we depict the real part of the \textit{PE} signal at the center ion of a three-ion chain coupled to a harmonic bath which is modeled by the Lindblad master equation
\begin{align}
\dot{\rho}(t) = -i[H_{\mathrm{ph}},\rho(t)] + \:&\gamma( \overline{n} + 1)\sum_{i=1}^N\left(a_i\rho(t) a_i^{\dagger}-\frac{1}{2}\{a_i^{\dagger}a_i,\rho(t)\}\right) \notag \\
+\:& \gamma\overline{n}\sum_{i=1}^N  \left(a^{\dagger}_i\rho(t) a_i-\frac{1}{2}\{a_ia^{\dagger}_i,\rho(t)\}\right). \label{eq.lindblad}
\end{align}
We assume local baths with identical temperatures corresponding to $\overline{n} = 0.1$. The two-exciton states are shifted by the anharmonic potential $U = - 0.03 \nu_x$. When the population time $t_2$ is set to zero [Fig.~\ref{fig.photon-echo} a)], we can observe six strong peaks.  They can be traced back to four near-harmonic transitions at $\Omega_{1} = 0.88\nu_x$ and $\Omega_2=1.00 \nu_x$, which can correspond to either $e-g$ transitions or unshifted $f-e$ transitions. Furthermore, two strongly anharmonic peaks, denoted $A$ and $B$ in Fig.~\ref{fig.photon-echo} a), show up at $\Omega_3 \approx 0.72\nu_x$ and $1. 1\nu_x$. The resonances are broadened by the coupling to the fluctuating environment.

To detect population decay, one has to monitor the photon echo signal as a function of $t_2$. Due to the coupling to the the heat bath, excitations can transfer from the environment to the system and vice versa. Since we chose a mean phonon number $\overline{n} = 0.1$ and weak coupling $\gamma = 0.0015\nu_x$ for all ions, populations predominantly leak from the system into the environment. 

Let us first discuss the effects of population decay at the hand of the diagrams in Fig.~\ref{fig.photon-echo-diagram}: The \textit{ESE}, \textit{ESAa} and \textit{ESAb} pathways evolve in a single exciton population during $t_2$, which can spontaneously decay to the ground state (or -- with much smaller probability -- get excited to a two-exciton state). When this happens, the signals from \textit{ESE} and \textit{ESAa} vanish, since the former one features a de-excitation of the ground state at the third interaction, and the latter one reduces to the ground state after the fourth interaction and does not fluoresce. When following the \textit{ESAb} pathway, the system can also decay spontaneously to the ground state during $t_2$, but the two successive pulses excite it back to a single exciton state, such that it still yields a signal. However, instead of a $f-e$ transition along $\Omega_3$, we expect to observe a $g-e$ transition. 

These effects are reflected in the photon echo signals in Figs.~\ref{fig.photon-echo} b) and c). With increasing $t_2$, the two anharmonic peaks at $\Omega_3 \approx 0.72\nu_x$ and $1. 1\nu_x$ lose intensity, and new peaks at $\Omega_3 \approx 0.85\nu_x$ emerge [denoted $C$ and $D$ in Fig.~\ref{fig.photon-echo} c)]. These effects can be attributed to the loss of the \textit{ESE} and \textit{ESAa} contributions, since the relative weight of the \text{GSB} and \textit{ESAb} pathways increases. The latter one is mostly responsible for the new peaks at $\Omega_3 \approx 0.85\nu_x$.

\subsection{Individual diagrams}
\label{sec.singlediagrams}
\begin{figure}
\centering
\includegraphics[width=.49\textwidth]{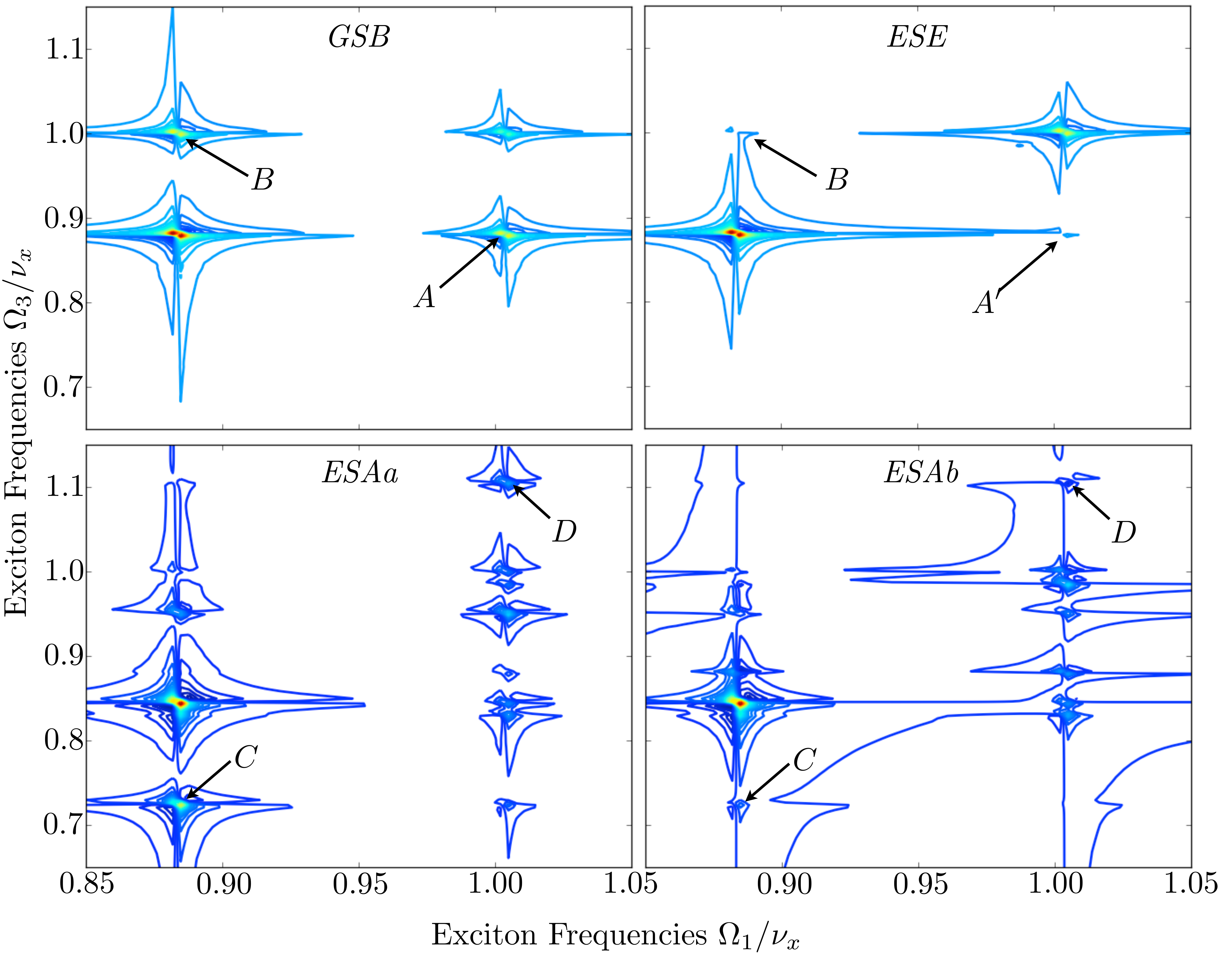}
\caption{(Color online) The real part of the diagrams comprising the  \textit{photon echo} signal $ S^{\textit{PE}}_{1111; 1} (\Omega_1, t_2, \Omega_3, t_4 = 0)$ for a thermal bath with temperature $\overline{n} = 0.1$, anharmonicity $U = -0.03 \nu_x$, and population time $t_2 = 0$. In contrast to spectroscopic experiments on molecular aggregates, we are able to investigate individual contributions to the full signal (see text). Peaks $A$ and $B$ point to harmonic crosspeaks, and peaks $C$ and $D$ highlight the strongly anharmonic $f - e$ transitions (see text).}
\label{fig.photon-echo2}
\end{figure}
As we noted in section \ref{sec.selectiveexcitation}, it is possible using strong resonant pulses to only trigger excitations or de-excitations of phonons. With regard to the \textit{PE} diagrams in Fig.~\ref{fig.photon-echo-diagram}, this provides the possibility to probe the individual diagrams that constitute the \textit{PE} signal. For instance, one can select the \textit{GSB} contribution using the sequence $V_+$, $V_-$, $V_+$, $V_+$ [cf. Eqs.~(\ref{eq.vplus}) and (\ref{eq.vminus})]. This is illustrated in Fig.~\ref{fig.photon-echo2}, where we depict the four diagrams of the photon echo separately for $t_2 = 0$ [see also the corresponding diagrams in Fig.~\ref{fig.photon-echo-diagram}]. Their sum yields the result in Fig.~\ref{fig.photon-echo} a). By decomposing the \textit{PE} signal into experimentally accessible, individual diagrams, we can observe that, both, the \textit{GSB} and the \textit{ESE} contribution, only show resonances at the single-exciton coherences. However, only the \textit{GSB} contribution shows crosspeaks between the single excitons (peaks $A$ and $B$). Furthermore, the two \textit{ESA}-pathways almost cancel each other in the sum due to their opposing relative signs, and only the anharmonic resonances (peaks $C$ and $D$) survive in the sum. Hence, by only monitoring for instance the \textit{ESAa} pathway, we can obtain ``background-free" signals, if we are mostly interested in higher-excited states. In particular, the strong resonance at $\Omega_3\approx0.85\nu_x$ can be observed in the individual signal, while it remains undetectable in the full photon echo signal, since it is canceled entirely by the \textit{ESAb} signal. The resolution of individual diagrams relies on the high flexibility for the design of interactions and thus represents a unique feature of cold-matter experiments, which is not within reach of current experiments on bulk materials or single molecules.

\section{Conclusions}
We presented a diagrammatic method to systematically construct nonlinear measurement protocols for synthetic quantum matter. This opens up a vast range of possibilities to analyze the dynamics of quantum optical systems of increasing complexity using tools from multidimensional spectroscopy. Moreover, we have discussed experimentally feasible implementations focussing on electronic and vibrational degrees of freedom in chains of cold trapped ions, making use of single-ion addressing. The latter property as well as the possibility to selectively create or annihilate excitations represent unique features of cold matter systems. These enable us to decompose the total signal into individual contributions which cannot be resolved in experiments where matter-field interactions are induced for instance by direct dipole transitions. We presented exemplary scenarios of two- and four-pulse measurements which allow to probe multi-point correlation functions of the underlying complex many-body dynamics.

Using two-pulse measurement protocols, we detected and analyzed steady-state currents in a chain of harmonic oscillators subject to a temperature gradient. Furthermore, we used the second-order signal to investigate the dynamics of excited states in a quantum Ising spin chain for different values of the relevant system parameters. When the system approaches a regime where two incompatible interactions compete, we identified signatures of chaotic excited state dynamics generated by superpositions of broadly distributed energies. This indicates a breaking of symmetry, which, on macroscopic scales, is signaled by a quantum phase transition. Four-pulse protocols allow for the investigation of transport of vibrational excitons beyond the manifold of single excited states, where anharmonicities in the trap potential can influence the effective coupling strengths between the ions. Specifically, we discussed the quantum optical analogues of spectroscopic signals, such as the \textit{double quantum coherence} and the \textit{photon echo}.

The present formalism may be applied similarly to other synthetic quantum systems. Even though single-site resolution is not necessary for the implementation of nonlinear measurement protocols according to the presented formalism, the most interesting applications are found when individual constituents of complex quantum systems can be controlled. In particular, we envision experimental realizations also with ultracold atoms in optical lattices and Rydberg atoms in optical tweezers, where current developments show promising prospects towards controlled coherent manipulation of single atoms in large interacting many-body systems \cite{Weitenberg11,Tweezer}.

This work shows that integration of concepts from nonlinear spectroscopy into quantum optical experiments establishes a promising line of research by combining two successful areas with backgrounds ranging from the very foundations of quantum physics to quantum chemistry.

\section*{Acknowlegments}
We thank M. Ramm and H. H\"affner for useful discussions and comments on the manuscript. F.S. and M.G. thank the German National Academic Foundation for support. S.M. gratefully acknowledges the support of the National Science Foundation through Grant No. CHE-1058791, and the Chemical Sciences, Geosciences and Biosciences Division, Office of Basic Energy Sciences, Office of Science, US Department of Energy.


\begin{thebibliography}{xx}
\bibitem{uncertaintyfootnote} The relative error for the difference of two signals is given by $\sqrt{2}s$, when $s$ is the relative error for the measurement of individual signals. Thus, a rough estimate for the error which allows to monitor such differences (for our set of parameters) is given by $s < 0.05/\sqrt{2}$.
\end{thebibliography}
\end{document}